\documentclass{aa}  

\usepackage{graphicx}
\usepackage{txfonts}
\newcommand\kms{$\mbox{km s}^{-1}$}

\newcommand\drthree{{\it Gaia}~DR3}
\newcommand\drfour{{\it Gaia}~DR4}

\begin{document}

   \title{Red supergiant stars in binary systems}

   \subtitle{II. Confirmation of B-type companions of red supergiants in the Small Magellanic Cloud using Hubble ultra-violet spectroscopy}

   \author{L. R. Patrick
          \inst{1},
          D. J. Lennon
          \inst{2},
		  A. Schootemeijer 
		  \inst{3, 4},
		  L. Bianchi 
		  \inst{5},
        I.~Negueruela,
		  \inst{6}
		  N.~Langer 
		  \inst{3, 4},
          D. Thilker
          \inst{3},
		  \and 
    		  R.~Dorda 
		  \inst{7}
          }

    \authorrunning{Patrick et al.}

   \institute{Centro de Astrobiologia (CAB), CSIC-INTA, Carretera de Ajalvir km4, 28850 Torrejón de Ardoz, Madrid, Spain\\
              \email{lrpatrick@cab.inta-csic.es}
\and
Instituto de Astrof\'isica de Canarias, E-38205 La Laguna, Tenerife, Spain\\
\and
Argelander-Institut f\"ur Astronomie, Universit\"at Bonn, Auf dem H\"ugel 71, 53121, Bonn, Germany\\
\and
Max-Planck-Institut für Radioastronomie, Auf dem H\"ugel 69, 53121, Bonn, Germany
\and
Department of Physics and Astronomy, Johns Hopkins University, 3400 N. Charles Street, Baltimore, MD 21218, USA\\
\and
Departamento de F\'{\i}sica Aplicada, Universidad de Alicante, E-03690 San Vicente del Raspeig, Alicante, Spain
\and
Universidad Europea de Canarias, La Orotava, Tenerife, Spain 
             }

   \date{Received September 15, 1996; accepted March 16, 1997}

 
  \abstract
  {Red supergiant stars (RSGs) represent the final evolutionary phase of the majority of massive stars and hold a unique role in testing the physics of stellar  models.  
  Eighty eight RSGs in the Small Magellanic Cloud (SMC) were recently found to have an ultra-violet excess that was attributed to a B-type companion.
  We present follow-up Hubble Space Telescope (HST) Space Telescope Imaging Spectrograph (STIS) ultra-violet (1700 -- 3000\,\AA) spectroscopy for 16 of these stars to investigate the nature of the UV excess and confirm the presence of a hot companion.
  In all cases we are able to confirm that the companion is a main-sequence B-type star based on the near-UV continuum.
  We determine effective temperatures, radii and luminosities from fitting the UV continuum with TLUSTY models and find stellar parameters in the expected range of SMC B-type stars. 
  We display these results on a Hertzsprung--Russell diagram and assess the previously determined stellar parameters using UV photometry alone.
  From this comparison we conclude that UV photometric surveys are vital to identify such companions
  and UV spectroscopy is similarly vital to characterise the hot companions.  
  From a comparison with IUE spectra of 32~Cyg, a well known RSG binary system in the Galaxy, four targets display evidence of being embedded in the wind of the RSG, like 32~Cyg, although none to the more extreme extent of VV~Cep. 
  The ages of six targets, determined via the stellar parameters of the hot companions, are found to be in tension with the ages determined for the RSG. 
  A solution to this problem could be binary mass-transfer or red straggler stars.  
  }  


   \maketitle
%

\section{Introduction}

Red supergiant (RSG) stars are the final evolutionary phase of the majority of massive stars before core collapse supernova.
Hydrogen rich supernova Type II-P are the most commonly observed type of core-collapse supernova and theory and observations agree that RSGs are most frequently the progenitor stars. 
The observed mass distribution of supernova Type II-P progenitors display a dearth of systems more massive than around 17-20\,M$_\odot$~\citep{Smartt09}.
Although see~\citet{2024arXiv241014027B} for a statistical assessment of this.
This can be interpreted as evidence that more massive stars transition through the RSG phase and explode at a different evolutionary phase.
The properties of observed populations of RSGs are sensitive tests to these evolutionary models.
Models predict that at certain masses and metallicities after the RSG phase stars evolve bluewards on the Hertzsprung--Russell diagram (HRD) and reside in the yellow or blue supergiant regime~\citep[YSG and BSG, respectively; e.g.][]{2012A&A...537A.146E,2019A&A...625A.132S}.
Indeed, there is some observational support for this based on mass-loss rates of YSGs~\citep[e.g.][]{2023arXiv230607336H}. 
However, recent results from the BLOeM campaign~\citep{BLOeM-I} in the Small Magellanic Cloud found that the multiplicity statistics of BAF-type supergiant stars are inconsistent with that of RSGs (Patrick et al. (accepted)).

The role of binary evolution in this picture is clearly significant but, for RSGs this is currently not well constrained.
Multiplicity among young massive stars is observed to almost unity~\citep[][]{Sana12,2014ApJS..213...34K,2017ApJS..230...15M} and the observed distributions of orbital parameters dictate that as stars evolve off the main sequence, they frequently exchange mass with a close companion~\citep{Sana12,2014ApJ...782....7D}. 
Such mass-transfer events are expected to result in stripped envelope stars in favour of RSGs~\citep{2008MNRAS.384.1109E}. 
Stars in binary systems that are close enough to merge produce rejuvenated projects that may subsequently evolve to the RSG phase~\citep[e.g.][]{2019Natur.574..211S}.
Such a merger is the current best explanation for the progenitor to the famous SN1987A~\citep{1992ApJ...391..246P,2017MNRAS.469.4649M} and play an important role in explaining the observed population of blue supergiants~\citep{menon2024}.
Evidence for these so-called `red straggler' stars in the RSG phase is best observed in the luminosity distribution of young massive star clusters (\citealt[][]{2019MNRAS.486..266B,b19,2020A&A...635A..29P}; Wang et al. in prep.), which is estimated to account for up to 50\,\% of the RSGs in clusters~\citep{b19,2020A&A...635A..29P}.

Identifying and characteristing the population of RSGs currently in binary systems is key to understanding the RSG population as a whole.
The known population of RSGs that are currently in binary systems is relatively small and poorly characterised in general with orbital periods known for only around 10 systems~\citep[][and references therein]{2024BSRSL..93..173P}.
Of these systems, the majority of companions are B-type main-sequence stars and the orbital periods range from around 1000\,d to 100\,yr.
These systems are frequently found to display evidence of interaction between the companions.
This typically assumed to be interaction between the wind of the RSG and the hot companion and the evidence for this is circumstellar emission line lines seen in the UV~\citep{1978A&A....70..227K}. 
In the case of the eclipsing binary VV~Cep~\citep{1977JRASC..71..152W} and KQ Pup the hot companions are embedded in a dense circumstellar material. 

The population of such well understood RSG binaries in the Milky Way is so small that it is unclear if selection biases play a role in their detection.
Despite this, this population serves to guide expectations. 
B-type main-sequence stars are the most common companion~\citep[see e.g.][]{2020RNAAS...4...12P} in the Milky Way population and other companions include a RSG-neutron star system~\citep{masetti06,2020ApJ...904..143H,2024BSRSL..93..173P} and a potential RSG+RSG system~\citep{2017A&A...606L...1W}.
From the early work of~\citet{burki} in the Milky Way, populations of RSG binary systems have been discovered in galaxies in the Local Group~\citep{2020ApJ...900..118N,2021ApJ...908...87N,2022MNRAS.513.5847P}
~\citep[see][for a review of such studies]{2024IAUS..361..279P}. 


In \citet[][; henceforth Paper~I]{2022MNRAS.513.5847P} we studied the ultra-violet (UV) properties of RSGs in the Small Magellanic Cloud (SMC) using UVIT~ASTROSAT photometry and discovered that 88 stars present a UV excess that we attributed to the presence of a binary companion. 
In this paper, the second in the series, we follow up a fraction of these companions with Hubble UV spectroscopy with the aim of confirming the nature of the targets and better characteristing the binary systems. 
The sample selection and observations are described in Section~\ref{sec:obs}. 
Section~\ref{sec:results} presents our key results, which we discuss in Section~\ref{sec:discussion}.
Finally, we conclude this article in Section~\ref{sec:conclusions}.

\section{Observations}      \label{sec:obs}
\subsection{Sample selection}       \label{sub:sample}

The sample has been selected to focus on the highest likelihood RSG binary systems in the SMC. 
All targets were required to have a clear UV-detection from the UVIT Imaging survey of the SMC (Thilker et al. in prep.). 
In this sense, the majority of the targets have been previously identified as RSG binary systems in Paper~I and have tentative stellar parameters determined using UV photometry alone. 
In addition, targets with radial velocity variability from~\citet[][; henceforth DP21]{2021MNRAS.502.4890D} were prioritised, which allows us to focus on the systems with the smallest separation. 

DP21 flagged LHA 115-S 7 as the most RV variable star in their sample and hence, a likely RSG binary system.
However, this source is not included in the study of \citet{2022MNRAS.513.5847P} as it is not in the RSG parent catalogue of~\citet{2019A&A...629A..91Y}.
By cross-matching the coordinates of the cool component with the UVIT SMC survey of Thilker et al (in prep.) we find that the UV-bright component to be almost one magnitude brighter in the FUV than any of the companions identified in Paper~I, therefore, we include this target in the sample for spectroscopic follow-up.
In total 28 targets were selected for follow-up observations.

\subsection{New Hubble STIS spectra}

As part of the Cycle 29 HST SNAP programme (16776; PI: L. Patrick), we observe STIS spectra of 16 RSGs with hot companions. 
The spectral types for the RSG primaries are listed in Table~\ref{tb:id}.
The \citet{1989AJ.....98..825S} KM-star catalogue of the SMC is used as the naming convention for sources in this article. 
The STIS/CCD G230LB configuration is used to obtain spectra covering the spectral range 1685–3065\,\AA\ with a spectral resolving power ($R$) of $\sim700$.
We use the 52x0.2E1 pseudo-aperture resulting in a 0.2" slit width at a favourable position on the detector.
Target acquisition focuses on the RSG star and the exposure time is determined on a star-by-star basis. 
Acquisition images use the F28X50LP aperture with the long-pass filter, which has a pivot wavelength of 7229\AA\ and full width at half-maximum (FWHM) of 2722\AA.
On visual inspection of these images, there are no other sources in any of the $5\times5"$ field of view images that may contaminate the near-UV spectra and each target is point-like and displays no obvious signs of blurring from a nearby source, which could be the source of the UV flux.
Despite the broad, red filter used in the acquisitions, tests with the HST exposure time calculator demonstrate that an early B-type star would be detectable at low-S/N in such acquisition images and their absence adds strength to the conclusion that source of the UV flux is coincident on the position of the RSG.
In addition, we cross-match each source with Gaia~DR3 and find no Gaia nearby sources that are bright enough to contaminate the UV spectra. 

\begin{table*}
\centering
\caption{Spectral types, IDs and stellar parameters of the RSG components of the systems.}
\label{tb:id}
\begin{tabular}{lllc ccccccc}
\hline
ID & ID-SIMBAD           & SpT & Ref.$^a$ & M$_{\rm RSG}$  & $\sigma$ & T$_{\rm eff}$ & $\log$L/L$_\odot$ & $\sigma$   & Age   & $\sigma$\\
& & & & \multicolumn{2}{c}{[M$_\odot$]} & [K] & \multicolumn{2}{c}{[$dex$]} & \multicolumn{2}{c}{[Myr]}\\
\hline
SkKM 74  & PMMR  44            &  K0.5Iab     & (1) & 13.2  & 1.745 & 4220  & 4.71  & 0.146 & 16.16 & 2.86 \\
SkKM 90  & Dachs SMC 1-13      &  K5Iab       & (2) & 16.8  & 2.203 & 4250  & 4.98  & 0.15  & 11.89 & 1.8 \\
SkKM 108 & SkKM 108            &  G8Iab       & (1) & 10.4  & 1.052 & 4440  & 4.43  & 0.141 & 22.46 & 3.5 \\
SkKM 113 & PMMR  60            &  G7.5Ia-Iab  & (1) & 11.5  & 1.374 & 4410  & 4.55  & 0.143 & 19.49 & 3.37 \\
SkKM 171 & PMMR  94            &  K1.5Iab-Ib  & (2) & 13.4  & 1.789 & 4130  & 4.73  & 0.146 & 15.8  & 2.8 \\
SkKM 209 & PMMR 118            &  K2Iab       & (2) & 13.3  & 1.764 & 4100  & 4.72  & 0.146 & 16.01 & 2.83 \\
SkKM 238 & SkKM 238            &  K0.5Iab     & (1) & 11.5  & 1.381 & 4260  & 4.56  & 0.143 & 19.44 & 3.37 \\
SkKM 239 & SkKM 239            &  K2Iab       & (2) & 14.1  & 1.95  & 4130  & 4.78  & 0.147 & 14.77 & 2.62 \\
SkKM 247 & Dachs SMC 2-10      &  K1.5Iab     & (1) & 14.1  & 1.949 & 4160  & 4.78  & 0.147 & 14.77 & 2.63 \\
SkKM 250 & RMC  29             &  K0Ia-Iab    & (2) & 14.4  & 2.019 & 4280  & 4.81  & 0.147 & 14.38 & 2.56 \\
SkKM 286 & BBB SMC 348         &  K0Ia-Iab    & (1) & 12.2  & 1.556 & 4210  & 4.62  & 0.144 & 18.06 & 3.23 \\
SkKM 310 & PMMR 178            &  G8-K2Ia-Iab & (1) & 14.4  & 2.014 & 4230  & 4.8   & 0.147 & 14.41 & 2.56 \\
SkKM 320 & BBB SMC 115         &  G8.5Ia-Iab  & (1) & 10.4  & 1.058 & 4360  & 4.43  & 0.141 & 22.39 & 3.5 \\
--       & LHA 115-S   7       &  G6.5Ia-Iab  & (1) & 11.0  & 1.234 & 4318  & 4.505 & 0.142 & 20.8  & 3.4 \\
--       & M2002 SMC  54134    &  G8Iab       & (1) & 11.3  & 1.317 & 4190  & 4.54  & 0.143 & 19.93 & 3.4 \\
--       & PMMR  61            &  K/M:        & (3) & 12.9  & 1.699 & 4130  & 4.69  & 0.145 & 16.59 & 2.95 \\
\hline
\end{tabular}
\tablefoot{$^a$ References for spectral types (1){ \citet{2015A&A...578A...3G}}, (2) {\citet{2018A&A...618A.137D}} and (3) {\cite{1983A&AS...53..255P}}}
\end{table*}

Each observation is a maximum of one full HST orbit and the signal-to-noise ratios (s/n) obtained range from 30 to 60 per resolution element.
Data reduction and spectral extraction is carried out using the latest version of the STIS official pipeline with the most recent version of the appropriate bad-pixel mask, which is only available around six months after observation, used in all cases.
An analysis of the reconstructed two-dimensional spectrum shows no evidence for any additional targets in the slit. 
SkKM~250 is the exception to this with a faint nearby target that does not contaminate the central spectrum.
We tested the scattered light correction of~\cite{2022AAS...24030203W} and find that the correction is on the $< 1.0\%$ level for our targets, but can reach up to $8\%$ in some cases.
Correcting for this effect does not alter the spectral appearance, nor any individual feature, but rather changes the continuum distribution, which may affect the model fitting analysis if left uncorrected.
Upon testing, it is clear that the scattered light correction is well within the uncertainties of the model fits.
    
In general, the overall spectral appearance of the targets are consistent with a normal B-type star with several interstellar lines, characteristic of the SMC, overlaid (more details are given in Section~\ref{sec:results}).
There are several examples of bad pixels which are not corrected by the final version of the bad-pixel mask for each observation, the most pronounced of which is at around 2900\AA, present in all of the spectra. 
We remove any remaining bad pixels on a case-by-case basis in the final spectra.





\subsection{UV photometry}

By design, the sample have UV magnitudes from the UVIT SMC survey (Thilker et al. in prep.) in the F172M filter with a limiting magnitude of 18.9\,AB, which is the limit to obtain sufficient S/N for our scientific requirements in one HST orbit.

To aid the characterisation of the hot companions we utilise point spread function fitting photometry as part of the Swift Ultraviolet Survey of the Magellanic Clouds~\citep[SUMaC;][]{2017MNRAS.466.4540H}.
SUMaC observed all targets in three photometric filters, namely, {\it uvw2, uvm2 and uvw1}, which have a PSF full width half maximum of 2.92, 2.45 and 2.37\arcsec\ and central wavelengths of 1928, 2246 and 2600\AA, respectively.
Further details of the survey are given in~\citet{2017MNRAS.466.4540H}.
Such photometric observations are an excellent complement to data from UVIT SMC survey.

By cross-matching our catalogue with the PSF photometry of~\citet{2017MNRAS.466.4540H}, we find that 14/16 of the observed sample have Swift photometry in all three photometric bands.
The two stars without Swift photometry (SkKM~310 and 320) 
fall outside the field of view of the SMC Swift survey.
The two wider Swift filters (\textit{uvw1, uvw2}) have a known red-leak, which results from the shape of the filter throughout profile and effectively allows bright red objects to contaminate the UV photometry. 
This is a particular consideration for our targets as, by construction, these sources have a bright RSG companion.
After tests with the photometry we only consider targets to have a meaningful detection of a binary companion if the source has a consistent detection in the \textit{uvm2} filter, which is sufficiently narrow that the filter does not suffer from the red leak problems.
All of the observed target have \textit{uvm2} data consistent with both the other \textit{Swift} filters and the UVIT detections from~\citet{2022MNRAS.513.5847P}.



   \begin{figure*}
   \centering
   \includegraphics[width=\linewidth]{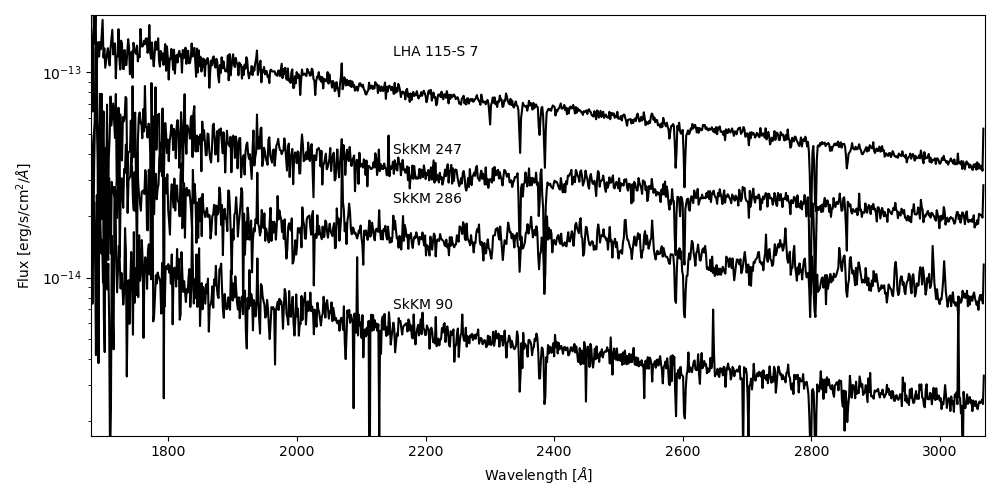}
   \caption{A montage of the observed HST/STIS UV spectra. Each spectrum has a signal-to-noise ratio per resolution element of at least 30 at $\sim$2400\,\AA, ranging to $\sim$60 for LHA~115-S~7. The flux distribution is not normalised to allow for better visualistion of the hot star continuum. No normalisation factor is applied to any of the spectra displayed here.}
              \label{fig:montage}%
    \end{figure*}

\begin{table*}[ht]
\caption{Summary of RV information for targets}
\label{tb:rv_info}
\begin{tabular}{l cccc ccc l}
\hline

& \multicolumn{4}{c}{DP21} & \multicolumn{3}{c}{\drthree}\\
ID                 &$\varv$ & Epochs  & $\Delta \varv$ & $\Delta$t & Epochs  & $\Delta \varv$ & $\Delta$t & Notes\\
                   &[\kms] &         & [\kms]         & [d]       &         &  [\kms] & [d] \\
\hline
SkKM 74               & 165.6\,$\pm$\,1.1   &  4  &  3.1  \,$\pm$\,1.4 &  11949   & 5 & 8.2  & 897 \\ 
SkKM 90               & 158.0\,$\pm$\,2.1   &  6  &  6.2  \,$\pm$\,1.3 &  5010    & 9 & 11.3 & 890 \\ 
SkKM 108              & 154.9\,$\pm$\,1.1   &  4  &  2.7  \,$\pm$\,2.6 &  2078    & 13 & 9.8  & 890 & Embedded \\ 
SkKM 113              & 143.18\,$\pm$\,0.94 &  3  &  2.2  \,$\pm$\,1.2 &  1082    & 10 & 7.8  & 971 \\ 
SkKM 171              & 177.51\,$\pm$\,0.85 &  7  &  2.3  \,$\pm$\,1.4 &  11567   & 9 & 7.6  & 946 \\ 
SkKM 209              & 148.6\,$\pm$\,8.8   &  7  &  23.3\,$\pm$\,1.0 &  11947   & 6 & 7.4  & 754 & Embedded \\ 
SkKM 239              & 153.60\,$\pm$\,0.84 &  6  &  2.3  \,$\pm$\,0.6  &  5010    & 4 & 7.5  & 908 \\ 
SkKM 250              & 182.0\,$\pm$\,3.0   &  8  &  9.6  \,$\pm$\,2.0 &  10572   & 5 & --   & 697 \\ 
SkKM 286              & 171.0\,$\pm$\,6.6   &  4  &  15.6\,$\pm$\,1.6 &  11537   & 7 & 10.6 & 890 & Embedded \\ 
SkKM 238              & 163.3\,$\pm$\,3.3   &  3  &  7.6  \,$\pm$\,1.2 &  1082    & 10 & 10.1 & 946 \\ 
SkKM 247              & 160.3\,$\pm$\,1.3   &  7  &  3.2  \,$\pm$\,1.3 &  11949   & 10 & 8.7  & 946 \\ 
SkKM 320              & 192.5\,$\pm$\,5.0   &  4  &  11.4\,$\pm$\,1.4 &  2078    & 2 & 11.9 & 873 \\ 
SkKM 310              & 196.4\,$\pm$\,2.4   &  7  &  6.8  \,$\pm$\,2.2 &  11919   & 3 & 6.0  & 873 & Embedded \\ 
LHA 115-S   7         & 102.1\,$\pm$\,17.2  & 4   & 39.2 \,$\pm$\,1.5&  2078    & 9 & 10.2 & 809 & H$\alpha$ emission\\ 
$[$M2002$]$~SMC~54134 & 159.1\,$\pm$\,10.7  &  2  &  21.4\,$\pm$\,1.4 &  --      & 8 & 13.6 & 908 \\ 
PMMR  61              & 162.9\,$\pm$\,9.2   &  3  &  19.5\,$\pm$\,2.0 &  11566   & 6 & 9.7  & 873 \\ 
\hline
\end{tabular}
\end{table*}

\subsection{Visual data}

As noted above, the targets have been selected for follow-up HST UV spectroscopy based on the availability of multi-epoch optical RV observations from DP21, which probe the orbital motion of the RSG. 
We cross match the 16 targets with \drthree\ finding good matches to within 1\arcsec\ for all targets. 
\drthree\ provides additional RV information for 15/16 of the targets observed with STIS.
A characterisation of the orbital configuration based on the RV curves of the systems remains, unfortunately, beyond the reach of currently available observations as a result of the combination of  sparse RV sampling and long suspected orbital periods. 

Table~\ref{tb:rv_info} provides a summary of the available number of RV information for each target.
A comparison of the peak-to-peak RV ($\Delta\varv$) from DP21 and that of \drthree\ shows that there is an inconsistency between the two measurements.
The \drthree\ $\Delta\varv$ measurement has no associated uncertainty but based on tests for the wider RSG population of the SMC, we assume that there is a systematic uncertainty associated with these measurements that require the time series RVs to resolve, which will be published in \drfour.
We retain this information in Table~\ref{tb:rv_info} to serve as a comparison, but do not consider it further in the text. 
That being said, the difference for LHA~115-S~7 is puzzling. 
One would assume that for large values of $\Delta\varv$ the Gaia data would be reliable.
Interestingly, if we exclude \drthree\ from the measurements of DP21 we find $\Delta\varv = 10$\,\kms,
which is consistent with the Gaia value. 
In addition to the systematic uncertainty discussed above, we consider it possible that there is a discrepancy between the Gaia and non-Gaia RVs for this source. 

In addition we compile photometric observations from the visual to the infrared from various sources to determine the spectral energy distribution (SED) of the RSG component and determine stellar parameters. 

Finally, none of the targets were classified as binary with the Gaia~DR3 NON\_SINGLE\_STAR flag. 
This flag identifies astrometric binaries and spectroscopic binaries, therefore, an expectation here may have been that some RSG binaries may have been picked up by the spectroscopic binary flag given that an excess is observed in the Gaia~XP spectra.

\section{Results}            \label{sec:results}


From the HST/STIS spectra all sources display a B-type star UV continuum. 
Based on this result, combined with the consistency of the positions of the UV source and the RSG component from~\citet{2022MNRAS.513.5847P} and the statistically insignificant possibility that a chance alignment occurs to such a high spatial accuracy~\citep[e.g.][]{2020ApJ...900..118N, 2022MNRAS.513.5847P,2024ApJS..274...45B}, we conclude that the stars form physically associated binary system. 
We assess the ages of the two components in Section~\ref{sec:ages}.
A montage of the spectra is shown in Figure~\ref{fig:montage}.
Four sources SkKM~108, 209, 286 and~320 show departures from a normal B-type star continuum, this is discussed further in Section~\ref{sub:interaction}. 

To test whether the RSG component contaminates the STIS spectra, we exploit the HST ASTRAL spectral database~\citep{astral} and rescale the spectra of $\alpha$~Ori (M2\,Ia) and $\beta$~Gem (K0\,III) to match the V-band magnitude of each target individually. 
From this comparison in no cases would a star like $\alpha$~Ori contaminate the spectra and only in three cases does the K0 star $\beta$~Gem have sufficient flux to potentially contaminate the near-UV spectrum: SkKM~171, 239 and 310. 
However, none of these targets, nor any others display clear signs of contamination from the RSG component. This is supported by the SED fitting analysis below.


\subsection{Stellar parameters from UV continuum fits}        \label{sub:ste_par}

To determine stellar parameters of the hot component we use model spectra from the TLUSTY SMC BSTAR and OSTAR grids with effective temperatures covering the range of 15,000 -- 55,000\,K~\citep{lanzhubenyO,lanzhubenyB}.
These spectra were convolved with the STIS G230LB line spread function and re-binned the STIS CCD pixel size. 
Given the modest resolution of these data, stellar rotation was ignored.
The low resolution of the data precludes the use of stellar absorption lines in determining spectral types, the strong lines that are present are dominated by their interstellar contribution.
The \ion{Fe}{iii} complex of lines in the vicinity of $\sim$1900\,\AA\ that is characteristic of early B-type stars, lies at the blue edge of the G230LB spectral range where the s/n is poor and hence not useful for constraining the effective temperature.
We therefore determined the stellar parameters by fitting the slope of the near-UV continuum.

Correction for extinction is a critical input to the fit process. 
However, experimenting with treating this as a variable \citep[cf][]{gordon2024} with just the near-UV bandpass proved unsuccessful.
\citet{2023ApJ...946...43W} provide individual values of extinction parameters for the stars in this sample but see the discussion on this in Appendix~\ref{sub:ebv}. 
We therefore opted to use a mean value of the extinction that we determined from a sample of 321 O-type stars with literature spectral types, taken mainly from \citet{2010AJ....140..416B}. 
O-type stars are ideal diagnostics of extinction as their $B-V$ intrinsic colour is almost degenerate with spectral type and metallicity~\citep{lanzhubenyO}.
The $B-V$ colours for the O-star sample were determined from their $Gaia$ XP spectra, and intrinsic colours were taken from~\citet{wegner1994} to infer a mean of $E(B-V)=0.08$ with standard deviation $0.06$. 
Following \citet{gordon2024} we assumed 0.04 magnitudes of foreground Milky Way extinction, and 0.04 magnitudes of SMC extinction, with appropriate average extinction laws.
The individual values of the $E(B-V)$ for the targets are listed in Table~\ref{tb:ebv} for comparison.

The TLUSTY model spectra spectra were reddened as above and a best fitting continuum was found using a chi-squared approach taking account of the uncertainties in the measured flux.
In general it was found that surface gravities in the range $\log\,{\rm g}=3.5-4.5$ gave similar quality fits, with lower surface gravities being significantly worse in this regard.
However, the G230LB data offer little leverage to refine this rough estimate and hence we do not report explicit values of log\,g other than to note that the sample is consistent with being on the main sequence. 
An additional caveat concerns the effective temperatures. For values close to, or above, 30\,kK the upper bound is poorly constrained as the continuum slope becomes insensitive at higher effective temperatures. 
At the cooler end of the scale we note that 15\,kK is the lower bound on the model grid, although the fits for the three stars with values near this edge appear reasonable (SkKM~239, 310 and 320).


\begin{table*}
\caption{Table of stellar parameters for the hot companions to RSGs as determined by the HST STIS Snaphot spectra. Ages are determined from comparisons with stellar models.}
\label{tb:params_m2}
\begin{tabular}{l ccccccc l}
\hline
ID & F172M & e$_{\rm F172M}$ & ${E(B-V)_{\rm Fit}}^a$ & T$_{\rm eff}$ & R & $\log L/L_\odot$ & Age & Notes\\
& & & & [kK] & [R$_\odot$] & [$dex$]& [Myr]\\
\hline
SkKM 74               &  14.95 & 0.04 & 0.08 & $25\,\pm3$ & $8.93 \,\pm\,0.5  $& $4.36\,\pm\,0.23$  & 18.38     & -- \\ 
SkKM 90               &  14.71 & 0.02 & 0.08 & $18\,\pm2$ & $18.9 \,\pm\,0.9  $& $4.44\,\pm\,0.22$  & 19.1     & -- \\ 
SkKM 108              &  15.80 & 0.03 & 0.08 & $27\,\pm3$ & $5.68 \,\pm\,0.17 $& $4.1 \,\pm\,0.2 $  & 16.79     & Embedded \\ 
SkKM 113              &  15.95 & 0.06 & 0.08 & $25\,\pm3$ & $6.12 \,\pm\,0.27 $& $4.04\,\pm\,0.22$  & 22.07     & -- \\ 
SkKM 171              &  18.73 & 0.13 & 0.08 & $18\,\pm2$ & $3.54 \,\pm\,0.18 $& $2.99\,\pm\,0.22$  & 87.97     & -- \\ 
SkKM 209              &  16.92 & 0.05 & 0.08 & $16\,\pm2$ & $7.78 \,\pm\,0.86 $& $3.47\,\pm\,0.27$  & 67.86     & Embedded \\ 
SkKM 238              &  16.04 & 0.04 & 0.08 & $27\,\pm3$ & $4.98 \,\pm\,0.17 $& $3.99\,\pm\,0.21$  & 10.84     & -- \\ 
SkKM 239              &  18.14 & 0.10 & 0.08 & $15\,\pm2$ & $5.1  \,\pm\,0.45 $& $2.99\,\pm\,0.25$  & 120.16     & -- \\ 
SkKM 247              &  16.56 & 0.07 & 0.08 & $30-40^b$  & $2.5-3.1$          & $3.87\,\pm\,0.23$  & 0.0     & -- \\ 
SkKM 250              &  15.44 & 0.08 & 0.08 & $30-40^b$  & $3.0-3.7$          & $3.92\,\pm\,0.22$  & 0.0     & -- \\ 
SkKM 286              &  15.57 & 0.05 & 0.08 & $23\,\pm2$ & $8.78 \,\pm\,0.6  $& $4.2 \,\pm\,0.23$  & 20.56     & Embedded \\ 
SkKM 310              &  18.86 & 0.21 & 0.08 & $15\,\pm2$ & $4.5  \,\pm\,0.37 $& $2.88\,\pm\,0.25$  & 117.18     & -- \\ 
SkKM 320              &  17.76 & 0.12 & 0.08 & $16\,\pm2$ & $6.32 \,\pm\,0.63 $& $3.29\,\pm\,0.27$  & 81.81     & Embedded \\ 
LHA 115-S   7         &  13.80 & 0.05 & 0.08 & $29-35^b$  & $11.0-12.7$        & $4.92\,\pm\,0.27$  & 9.97     & H$\alpha$ emission \\ 
$[$M2002$]$~SMC~54134 &  15.54 & 0.03 & 0.08 & $25\,\pm3$ & $7.4  \,\pm\,0.2  $& $4.2 \,\pm\,0.2 $  & 20.23     & -- \\ 
PMMR  61              &  15.66 & 0.04 & 0.08 & $21\,\pm2$ & $9.53 \,\pm\,0.51 $& $4.12\,\pm\,0.22$  & 24.8     & -- \\ 
\hline
\end{tabular}
\tablefoot{$^a$ Fit parameters are determined assuming an SMC average $E(B-V) = 0.08$. For reference we also list the $E(B-V)$ values found for nearest 20 O-type stars in the Table~\ref{tb:ebv}.
           $^b$ For these targets we provide a range of parameters that are consistent with the observations. See text for more details. 
}
\end{table*}

\begin{figure*}
   \centering
   \includegraphics[width=0.45\linewidth]{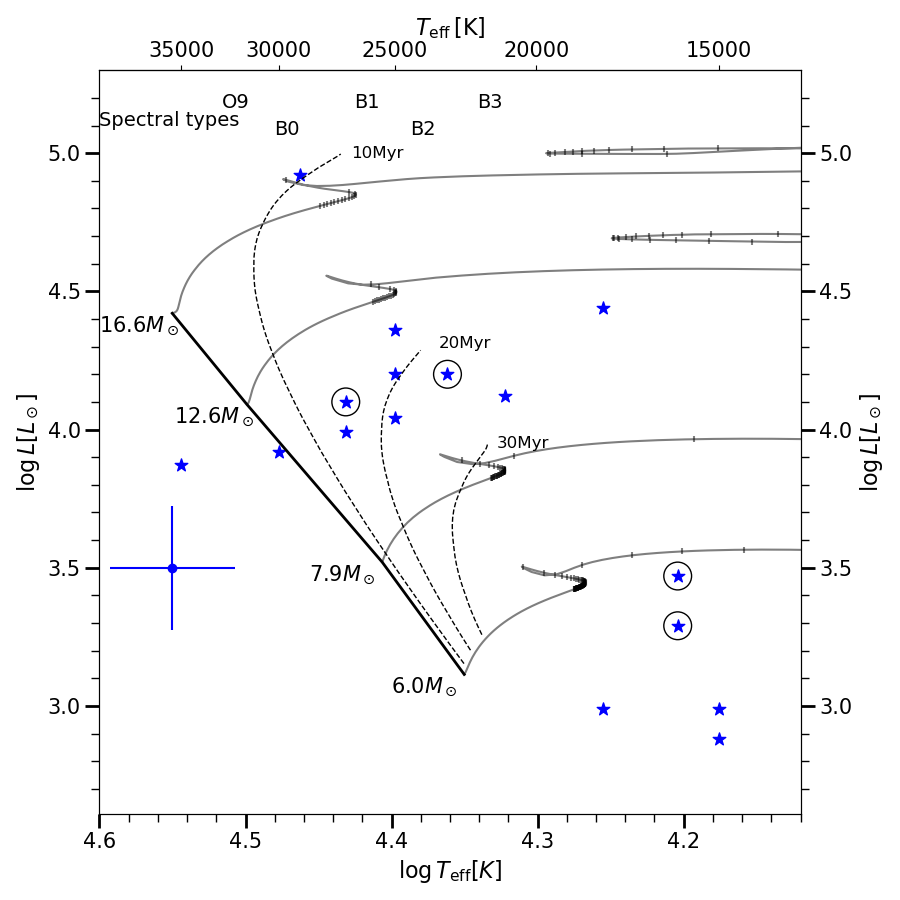}
   \includegraphics[width=0.45\linewidth]{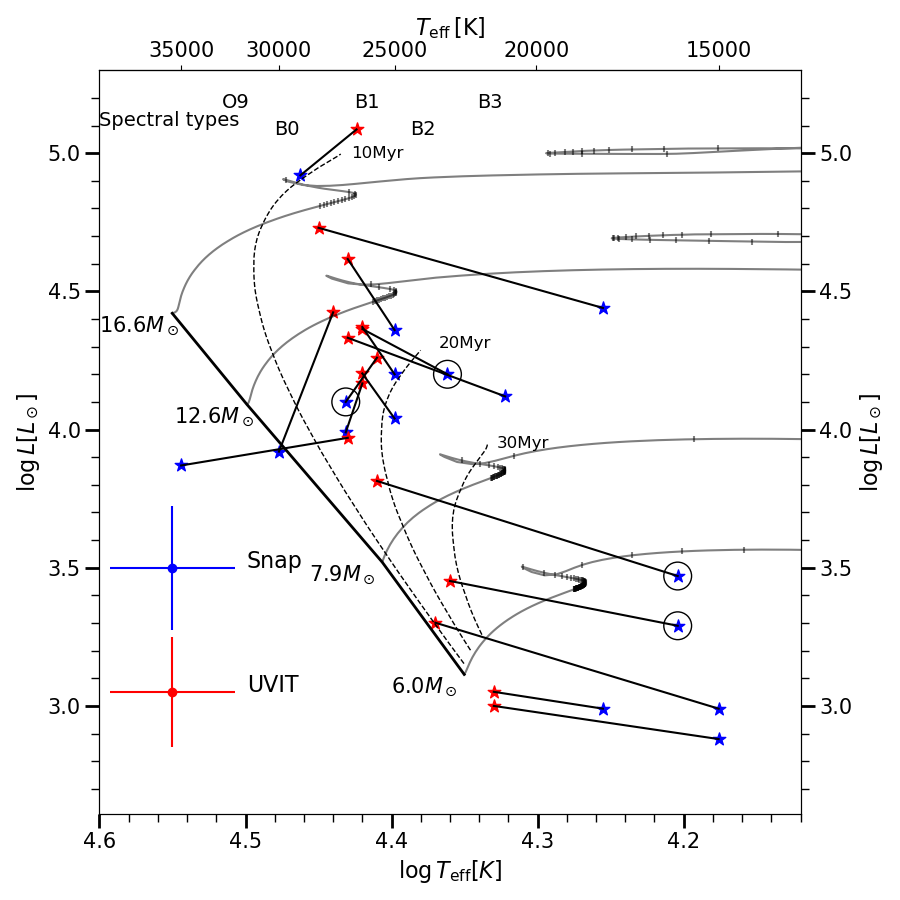}
   \caption{\textbf{Left:} Hertzsprung--Russell diagram (HRD) for the 16 targets with stellar parameters determined from the G230LB STIS spectra shown with blue stars. 
   Grey lines show SMC stellar tracks using a mass-dependent overshooting parameter~\citep{2019A&A...625A.132S,2021A&A...653A.144H}, where dashes are marked at intervals of 0.05\,Myr.
   The dashed lines show isochrones of ages 10, 20 and 30\,Myr. 
   Spectral types for SMC dwarf stars are shown using the calibration presented in~\citet{BLOeM-I}. 
   Large black circles mark the stars labelled as `embedded' in Table~\ref{tb:params_m2}.
   \textbf{Right:} Same as left panel but in this figure we add the stellar parameters determined from~\citet{2022MNRAS.513.5847P} in red. Black lines connect the stellar parameters derived using the two methods.
   } \label{fig:HRD}
    \end{figure*}

The results of this fitting using a fixed $E(B-V)= 0.08$ are shown in Table~\ref{tb:params_m2} for each source as well as their UVIT magnitudes.
In addition, Figure~\ref{fig:HRD} displays the stellar parameters graphically on a HRD, which also shows stellar tracks of different ages.
To assess the uncertainties on the fits we repeat this fitting process with 
a different model grid, that of the ATLAS9 models of~\citep{howarth2011} with a range of different assumptions.
The uncertainties listed in Table~\ref{tb:params_m2} reflect the distribution of resulting parameters.
In most cases the fits are robust to varying $\log {\rm g}$ in the range $3.0 < \log {\rm g} < 5.0$ and assuming a range of $E(B-V)$ values.
This includes the  sources at the low edge of the effective temperature grid and in addition, we find no evidence that the four stars labelled as `embedded' in Section~\ref{sub:interaction} have poorer fits.

The effective temperatures span the full range of the grid from $15 - 35$\,kK with luminosities in the range $3 < \log {L/L}_\odot < 5$. 
LHA~115-S~7, has the most luminous companion. Excluding this star the luminosity distribution peaks at $\log {L/L}_\odot = 4.4$.
This distribution of luminosities and effective temperatures is in good agreement with the known population of B-type main-sequence stars in the SMC.  
When displayed on the HRD in Figure~\ref{fig:HRD}, there appears to be two groups of targets, a low-luminosity group with effective temperatures less than 18\,kK and luminosities less than $\log L/L_\odot = 3.5$ and a higher-luminosity group with $ 4.0 < \log L/L_\odot < 4.5$. 
Such a distinction is unexpected given the roughly continuous distributions of UVIT~F172M magnitudes, luminosities and effective temperature of the companions although small number statistics prevents any conclusions to be drawn from such an apparent bimodal distribution. 

\subsection{Spectral energy distributions} \label{sub:SED}
We determine effective temperatures and radii of the RSG component from SED fitting of photometry, assuming a distance modulus to the SMC ($dm_{\rm SMC} = 18.977$; \citealt{2020ApJ...904...13G}). 
Figure~\ref{fig:sed_rsg} illustrates an example fit.
An examination of these fits, which are shown in Appendix~\ref{sec:SEDs}, reveals that in all cases the flux contribution from the RSG component is negligible in the near-UV spectral range, in good agreement with the ASTRAL analysis. 
The photometry compiled is from Gaia~DR3 and 2MASS data~\citep{2006AJ....131.1163S} and in the fitting procedure the photometry is weighted by their respective uncertainties.
In addition, these figures reveal that the nature of the hot companion is uniquely revealed by the near-UV spectra. The $U$-band in some cases shows evidence for contamination from the hot companion, but is in all cases fit reasonably well with a single RSG SED. 

   \begin{figure}
   \centering
   \includegraphics[trim=0cm 0cm 0cm 0cm, clip,width=\linewidth]{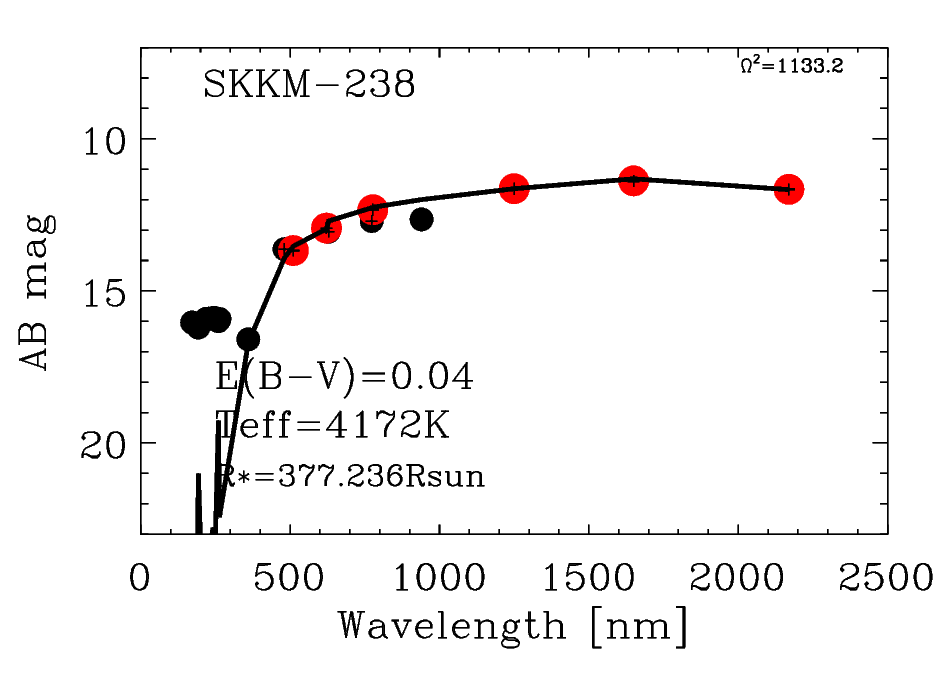}
   \caption{An example of the SED fitting to determine the stellar parameters of the RSG component.
            The large red points are from Gaia and 2MASS and are the only photometry given weight in the fit. Black points are from Skymapper, Swift and UVIT.}
              \label{fig:sed_rsg}
    \end{figure}

Stellar parameters are determined using the best fit Castelli \& Kurucz model atmospheres to the V- through K-band photometry assuming a fixed surface gravity of $\log g = 1.0$ and metallicity of $Z = 0.002$ or 14\,\% solar, which is typical of the SMC, and an SMC-like extinction curve. 
We compare the stellar parameters from the SED fitting and find very good agreement in general.
For all stars apart from LHA\,115-S~7, the agreement is well within assumed uncertainties on the individual parameters. 
The median difference between the effective temperatures is $\pm60$\,K and $\pm0.03\,dex$ in luminosity. 
In general, the fitting process prefers very low reddening values, which leads to $E(B-V) = 0.0$ in some cases. Repeating this process with a different extinction law improves the fit in some cases but the reddening values remain small. 
The change to the stellar parameters is minimal.
From this analysis we conclude that the $E(B-V)$ values are relatively poorly constrained from our fits, but low $E(B-V)$ are in general preferred. 
Comparing these results to the adopted $E(B-V)$ values from the OB-type star comparison shows no obvious trend.


The stellar parameters for LHA\,115-S~7 are not well fit, which is likely the result of the prominent H$\alpha$ emission feature perturbing the fit.
Repeating the fits using different reddening, $\log g$ and metallicity values illustrate that the stellar and extinction parameters are poorly constrained for this source. 
Stellar parameters determined from APOGEE data for this star~\citep{2022ApJS..259...35A} result in T$_{\rm eff} = 4278$\,K, which is in good agreement with the effective temperature determined using the analysis of Paper~I.
However, the $E(B-V)$ value from APOGEE is considerably larger than the values that the value returned by the fitting routine and from the OB-star analysis.

\section{Discussion}            \label{sec:discussion}
\subsection{Comparison with stellar parameters from single epoch photometry }
        \label{sub:comparison}
In Paper~I we determined stellar parameters for the hot companions by making three simplifying assumptions, the first is that the hot star is a main-sequence star, the second is that the age of the hot companion is the same as the RSG and is determined by that of the RSG and the third is the extinction parameters are same for both components. 
With the results presented for the HST near-UV spectra in Section~\ref{sec:results}, we are able to assess some of these degeneracies and improve upon the earlier measurements, both in terms of accuracy and precision.
Figure~\ref{fig:HRD} shows a direct comparison between the stellar parameters determined in Paper~I and those determined here.
In general, the effective temperatures are significantly lower than those determined in Paper~I, often by several thousand Kelvin. 
This is not unexpected given the assumptions made in Paper~I.
Coupled with this, the luminosities determined are consequently smaller using the HST/STIS spectra.
Here the difference is less significant given the large uncertainties.\footnote{The median uncertainties of the luminosities of 16 targets from Paper~I is $\pm0.2$\,dex, which is larger than their typical mean uncertainties of the whole sample.}


In 10/16 stars the effective temperature is significantly smaller than determined in Paper~I and in all cases the luminosity determined is smaller than those in Paper~I.
For the six stars that have higher effective temperatures from the HST/STIS spectra, the difference is significant only for the two hottest stars where it is likely that the uncertainties are underestimated. 
To test whether the difference in stellar parameters is the result of a discrepancy between the HST spectra and the UVIT photometry we compute synthetic photometric in the UVIT FUV band from the HST/STIS spectra.
On average the UVIT observations are consistent with the HST/STIS spectra to within a standard deviation of $\pm0.4$\,mag, which is larger than expected.
The origin of such large discrepancies is unclear and may be the result of inaccurate absolute flux calibration of the UVIT observations or potentially inaccuracies in the centering of the target on the slit in the HST/STIS spectra.
If we recompute the stellar parameters using the synthetic UVIT observations with the same routines as in Paper~I we find that the effective temperatures are always within the uncertainties but are usually between 0 and 400\,K warmer.
Similarly, the luminosities are typically within $\pm0.15\,dex$ in all but two cases.
Based on this analysis we rule out that the differences between the determined stellar parameters from the UVIT and HST data are the result of differences in the flux calibration between the two data sets.

\subsection{Are the ages of both components consistent?}            \label{sec:ages}
In Paper~I we concluded that 15/16 of the targets in this study had coeval components.
LHA\,115-S~7 was not in the sample of stars studied in Paper~I, but by determining the stellar parameters for both components using the same analysis routines, we find that this system would have been added to the six systems identified that had $q > 1$.
Perhaps the most important consequence of revising the stellar parameters of these stars is the realisation that the hot companions are, in general, significantly older and less massive than assumed in Paper~I.

In Paper~I we determined ages for the RSG component via a comparison with the same stellar evolutionary models as used in this study.
To do this we determined a calibration between the RSG luminosity and the mid-point of the core-helium burning sequence in the models.
While this assumption is simplistic, at 8\,M$_\odot$ the star spends only 1.6\,Myr in the RSG phase and throughout this time the luminosity is roughly constant.
This illustrates that even if the RSG is more or less evolved than expected, the determined age would not be significantly affected. 
Based on this, we assume that this assumption provides a realistic estimate of the age of star, although this neglects the fact that at the very end of their lifetimes RSG ages are very uncertain~\citep[e.g.][]{2020MNRAS.494L..53F}. 

\begin{figure*}
   \centering
   \includegraphics[width=\linewidth]{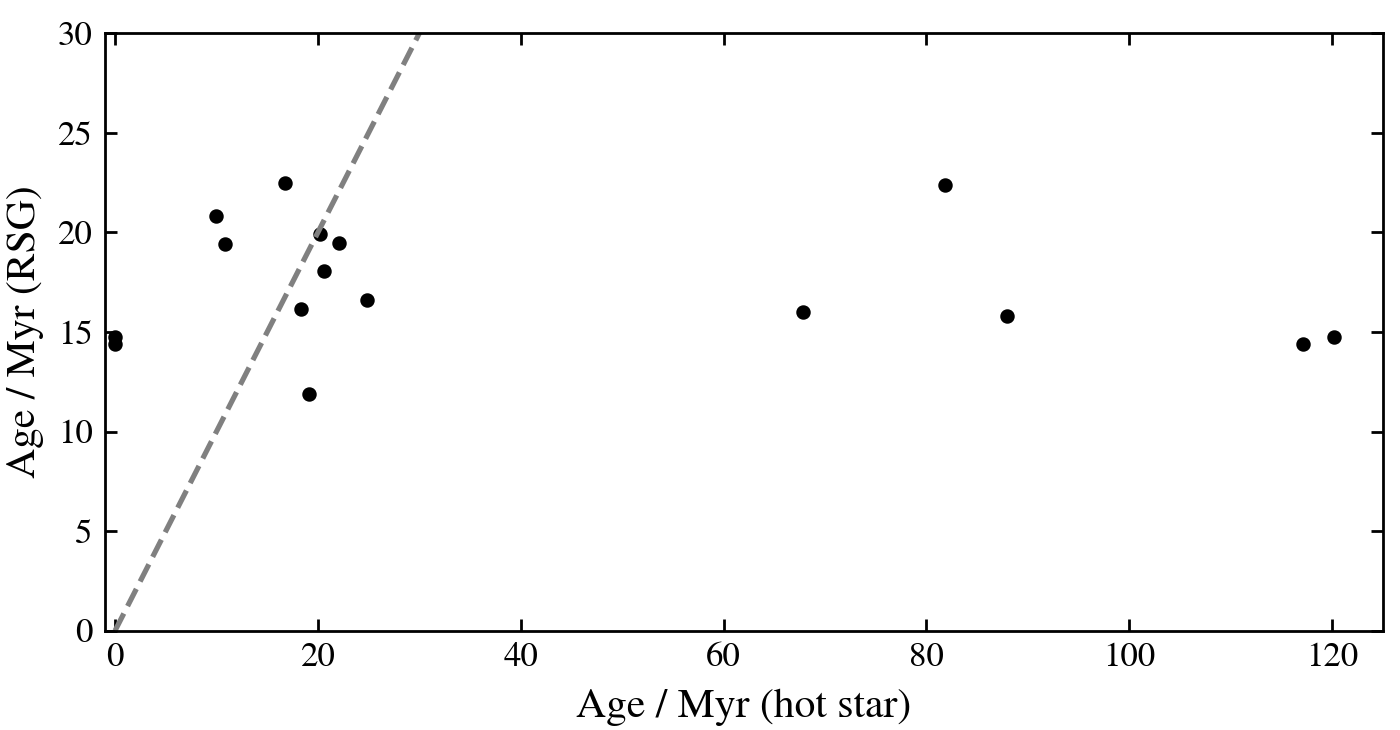}
   \caption{Comparison of the ages of the hot companions (abscissa) and the RSG components (ordinate) of the 16 binary systems studied. Black dashed line shows is the expectation of co-evality i.e. the one-to-one. The ages of both components are inconsistent in 7/16 cases.}
              \label{fig:age}%
    \end{figure*}

Ages for the hot companions are determined by matching their determined stellar parameters to the models.
For these measurements we assume uncertainties of $\pm5$\,Myr as a conservative estimate to take into account the uncertainity in $\log {\rm g}$. 
Figure~\ref{fig:age} compares the ages determined from the stellar parameters of the hot companions with those from the RSGs from Paper~I.
The ages for the 16 RSGs range between $12-22$\,Myr with uncertainties between $\pm$2 and 3.5\,Myr, which increases as a function of the determined age.
In contrast, the ages of the hot companions are $0-120$\,Myr. 
The majority of the hot companions show ages broadly consistent with that of the RSGs.
However, for five sources the ages of the RSG is in clear tension with that of the hot companions. 
We experiment with different stellar models from~\cite{2019A&A...625A.132S} assuming different mixing parameters with convective overshooting ranging between 0.11 and 0.55.
A larger convective overshooting parameter extends the main-sequence to cooler effective temperatures but does not relieve the tension between the ages of the two components.
For example, a seven solar mass model spend the first 40\,Myr the very close the position of the zero age main-sequence regardless of the assumed mixing parameters. 

Our conclusion is that for 6/16 sources some form of binary evolution must have taken place in order to explain the apparently different ages of the two components. 
An alternative explanation to explain the age discrepancy would be that the stars are not physically associated with each other.
Given that two of the six hot companions show some evidence of interaction (see below) this is clearly not the case for these sources and our interpretation is that chance alignment is unlikely to be able to explain all or indeed any of these sources. 
Interestingly, to the authors knowledge, there are no cases of a tension between the ages of different components in the best known Galactic population of RSG binary systems.

For the older B-type stars, an explanation for the age discrepancy is the so-called `red straggler stars'~\citep{b19}, which are the RSG counterparts of blue straggler stars. 
This would imply that these systems were originally in hierarchical triple systems where the inner binary has interacted, merged and subsequently evolved to the RSG phase. 


\subsection{Evidence of interactions} \label{sub:interaction}
\begin{figure*}
   \centering
   \includegraphics[width=\linewidth]{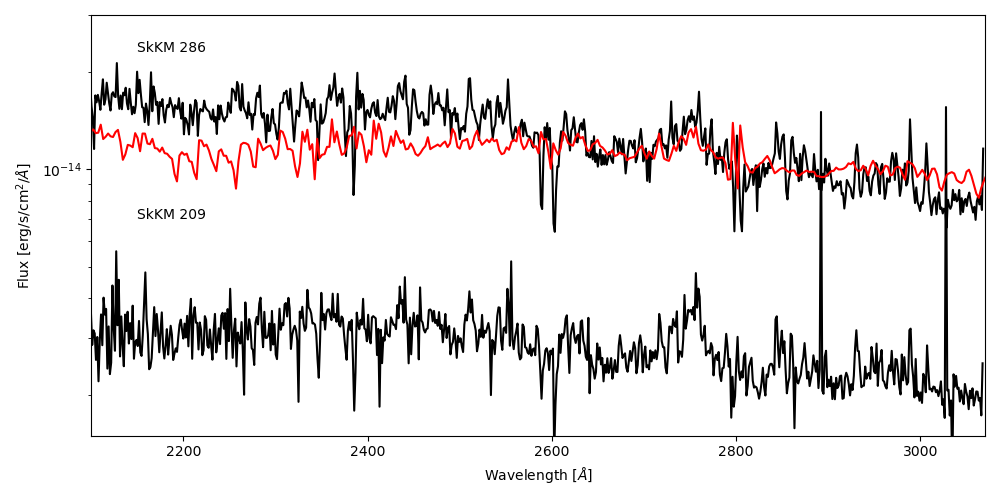}
   \caption{Two examples of HST/STIS G230L spectra of displaying deviations from a normal B-type star continuum. This behaviour is observed in SkKM~108, 209, 286 and 320 (209 and 286 shown in this figure). The red spectrum shows an IUE spectrum of 32~Cyg published in~\cite{1983A&A...126..225C} at an orbital phase of 0.7.}
              \label{fig:wiggles}%
    \end{figure*}

Four targets stand out as displaying an apparent second component in the HST/STIS spectra.
This manifests as broad emission features in the near-UV continuum, the strongest of these at $\sim$2750\,\AA.
The targets that display such features are SkKM~108, 209, 286 and~320. 
Figure~\ref{fig:wiggles} show SkKM~209 and 286 to illustrate this effect. 
This behaviour is not expected in normal B-type stars. 

From a comparison with stars in the ASTRAL database in Section~\ref{sec:results} and via the SED fitting presented in Section~\ref{sub:SED} we show that the RSG does not contaminate the near-UV continuum in any cases, even for the mid-G-type supergiant LHA\,115-S~7. 
We therefore exclude this as the origin of the second component in the spectra.  

To attempt to understand the origin of these spectral features we compare the observed spectra with archival International Ultra-violet Explorer (IUE) observations.
We compare the HST/STIS spectra with various RSG binary systems in the Galaxy with main-sequence B-type star companions~\citep[see][for a recent summary of the orbital parameters of such systems]{2024BSRSL..93..173P}, as well as to single B0V -- B3V stars. 
To do this we use the IUE INES archive and download spectra at various orbital phases for VV~Cep, 22~Vul, 31~Cyg, 32~Cyg, HR6902 and $\zeta$~Aur. 
These spectra were then degraded to the resolving power of the G230L spectra to aid a visual comparison to the observed features in the HST/STIS spectra.
From these comparisons we find that three spectra of 32~Cyg provide the best match.
These spectra were observed on 30/01/1980, 04/05/1980 and 01/08/1980 at an orbital phase of 0.62, 0.7 and 0.78, respectively
\citep{1983A&A...126..225C}, with the spectrum at the orbital phase of 0.7 matching the most closely.
Spectra of 32~Cyg at other orbital phases do not show the same features observed in the HST/STIS spectra, rather the spectra more resemble that of normal B-type stars.  
There are some similarities in the shape of the emission feature at 2750\,\AA\ with the IUE spectra of VV~Cep at 04/10/97, 16/06/97, 21/08/97 but here the agreement is clearly poorer in particular with in the Mg\,{\sc II} lines being blended into a strong emission profile in the VV~Cep spectra and a general lack of agreement with the shape of the near-UV continuum.

Figure~\ref{fig:wiggles} also shows the closest matching IUE spectra of 32~Cyg are observed on the 04/05/1980. 
We argue that this comparison shows that the spectra of these four stars can potentially be explained by the same mechanism that produces the spectrum of 32 Cyg. 
\citet{1979ApJ...233..621S} first identified P~Cygni line profiles in the Fe\,II and Mg\,II lines (among other elements), which is evidence that the hot star is embedded in the wind of the cool supergiant.
Such P Cygni profiles can explain the lack of strong absorption in the Mg\,II interstellar lines, which is also weaker in these four sources.
The hot companion of VV~Cep is also embedded in the wind of the RSG, in this system, however, the emission features that dominate the spectral appearance are far stronger and bear little resemblance to what is observed in the SMC systems, which perhaps is to be expected given the significantly later spectral appearance of the RSG than the stars studied here. 
High resolution UV spectra at multiple orbital phases is required to further this analysis and in a third paper in this series we will assess this further.
The four targets that display this behaviour are marked as `embedded' in Table~\ref{tb:params_m2}.

As additional supporting evidence to the embedded interpretation, 3/4 of the embedded targets have large observed $\Delta\varv$ values, which is indicative of orbital motion.
SkKM~108 shows no evidence of orbital motion which may indicate a significantly larger orbital period. 
Similar embedded spectral features are observed in the IUE spectra of $\alpha$~Sco, which has an orbital period of several thousand years and consequently the orbital motion displayed in this system is also very small. 

Eight of the the remaining objects show no indication of orbital motion and do not display evidence of interactions.
This could indicate significantly larger orbital periods.
However, four targets do show evidence of $\Delta\varv$ values that are larger than expected for normal intrinsic RSG variability:
LHA~115-S~7, $[$M2002$]$~SMC~54134 and PMMR~61. 
The strong H$\alpha$ emission line profile in LHA~115-S~7 is evidence for interaction, but no evidence is seen of this in the near-UV continuum as might be expected.
The remaining two stars ($[$M2002$]$~SMC~54134 and PMMR~61) appear to show no evidence for interaction but exhibit orbital periods short enough to display significant orbital motion.


\subsection{Discussion of individual targets} \label{sub:individuals}

\subsubsection{SkKM 250}
The 3-$\sigma$ significant Gaia parallax of SkKM~250 implies a geometric distance of 12$^{+7}_{-4}$\,kpc~\citep{2021AJ....161..147B}.
The kinematic properties of this source all appear consistent with SMC membership and various luminosity classifications place this star in the SMC. However, if the metallicity of the star is significantly lower than assumed when determining the spectral type, the luminosity class is likely to be unreliable. 
In their analysis of SMC RSGs, based on the elemental abundance pattern compared to the rest of their sample, \citet{1997A&A...324..435H} considered whether this star is a galactic halo star. 
These authors concluded that this star was a member of the SMC.
Using the neural network classifier of~\citet{2023A&A...672A..65J.SMC} the likelihood of SMC membership for SkKM~250, based on its Gaia~DR3 measurements is 51\,\%, which would mean that this source would be included in both their `complete' and `optimal' samples. 
The RV of this star is also fully consistent with SMC membership and there are no nearby Gaia sources which may confuse classifications.

If we assume that the distance provided by~\citet{2021AJ....161..147B} is correct this implies that SkKM~250 is giant star in the halo of the MW. The luminosity of the hot companion in this scenario would be $\log {\rm L/L}_\odot = 2.52\pm0.40$, assuming the 16th to 84th percentile confidence interval on the~\citet{2021AJ....161..147B} distance to determine the uncertainties on $\log$L.
This would make the hot companion a subdwarf OB-type star. 
Typically sdB stars have luminosities up to $\log {\rm L/L_\odot}\sim 1.5$~\citep{1997A&A...319..109M} and $\log {\rm L/L_\odot} = 2.52\pm0.40$ is potentially larger than theoretically possible for such stars~\citep{2002MNRAS.336..449H}. Therefore, we prefer the explanation that SkKM~250 is an SMC member, but acknowledge that the distance uncertainty makes the interpretation as a Galactic halo star a viable option. We assume the Gaia~DR4 parallax will resolve this uncertainty, but if not, a full binary solution may be the best method to distinguish between these scenarios.
We retain the in the sample and consider it an SMC member in the discussion.


\subsubsection{SkKM 247}
SkKM 247 has among the highest temperature determined in the sample, which for the luminosity of the star, places it far to the right of the main-sequence.
The effective temperature range for this star is 30--40\,kK, with a radius that appears smaller than that of the other sources.
To make this source consistent with the main-sequence position of the models requires a temperature at the lowest end of this range, this in turn implies very small, if not negligible, reddening values. 
In contrast, the SED fitting this source has among the largest determined $E(B-V) = 0.11$, which is still relatively modest. 
In addition, as noted in Appendix~\ref{sub:ebv}, this source shows a $K_{\rm s} - [24\mu{\rm m}]$ excess that suggests a reddening value of $E(B-V) = 0.096$, in good agreement with the SED fitting.
This is marginally larger, but fully consistent with the value determined from the comparison with nearby O-type stars (i.e. $E(B-V) = 0.076$). 
Assuming $E(B-V)=0.11$ from the SED fitting, 
places the effective temperature around 40\,kK, clearly to the left of the main-sequence given its luminosity. 
The Swift UVM2 absolute magnitude of this source is comparable with a $\sim$5\,M$_\odot$ B-type SMC stripped star~\citep{2023Sci...382.1287D},
which would imply an effective temperature of at least 60\,kK, which is inconsistent with the temperature estimates from the near-UV spectra.
In addition, at this mass we might expect to observe He\,{\sc II} absorption features~\citep{2018A&A...615A..78G} which are absent from the spectrum.

At present we consider this source as broadly consistent with being on the main-sequence but consider this source the best candidate so far for a stripped star companion to a RSG. 
This will be assessed in greater detail in a third paper in this series using shorter wavelength HST spectra.

\section{Conclusions} \label{sec:conclusions}

\begin{itemize}
    \item From near-UV HST/STIS spectral appearance all the hot companions are normal B-type main-sequence stars, in good agreement with expectations.
    \item Stellar parameters are well determined from the current spectra. Using TLUSTY model fits to the near-UV spectra, stellar parameters show considerable variation with those determined from UV-photometry alone. In general, the companions are cooler and subsequently less luminous that determined in Paper~I. 
    \item At least four systems show evidence of interaction, which -- at the resolution of the HST/STIS spectra -- manifest as broad emission features that deviate from the continuum expected from normal B-type stars. 
    \item Follow-up far-UV spectra will improve the accuracy of the stellar parameters for 11 of these targets including two of the interacting systems. In addition, follow up multi-epoch spectroscopic observations from the 4MOST One thousand and one Magellanic Cloud survey combined with Gaia~DR4 and the long baseline information from DP21, may provide some constraints on orbital periods. 
\end{itemize}

\begin{acknowledgements}
The authors thank Dr F. Najarro and Dr M. Garcia for helpful discussions and advice on IUE spectra. 
The authors thank Prof. M. Siegel for providing us with the \textit{Swift} photometry.
LRP acknowledges support by grants
PID2019-105552RB-C41 and PID2022-137779OB-C41 funded by
MCIN/AEI/10.13039/501100011033 by "ERDF A way of making
Europe". 
LRP acknowledges support from grant PID2022-140483NB-C22 funded by MCIN/AEI/10.13039/501100011033.
IN is supported by the Spanish Government Ministerio de Ciencia, Innovaci\'on y Universidades and Agencia Estatal de Investigación (MCIU/AEI/10.130 39/501 100 011 033/FEDER, UE) under grants PID2021-122397NB-C21/C22. It is also supported by MCIU with funding from the European Union NextGenerationEU and Generalitat Valenciana in the call Programa de Planes Complementarios de I+D+i (PRTR 2022), project HIAMAS, reference ASFAE/2022/017.
Contributions by LB and DT to this work were supported by grant HST-GO-16776 from the Space Telescope Science Institute under NASA contract NAS5-26555.
Based on observations with the NASA/ESA Hubble Space Telescope
obtained at the Space Telescope Science Institute, which is operated by the Association of Universities for
Research in Astronomy, Incorporated, under NASA contract NAS5-
26555.
\end{acknowledgements}

%
\bibliographystyle{aa} 
\bibliography{refs} 
%






   
  



\begin{appendix}

\section{HST spectra}
\begin{figure*}
   \centering
    \includegraphics[width=\linewidth]{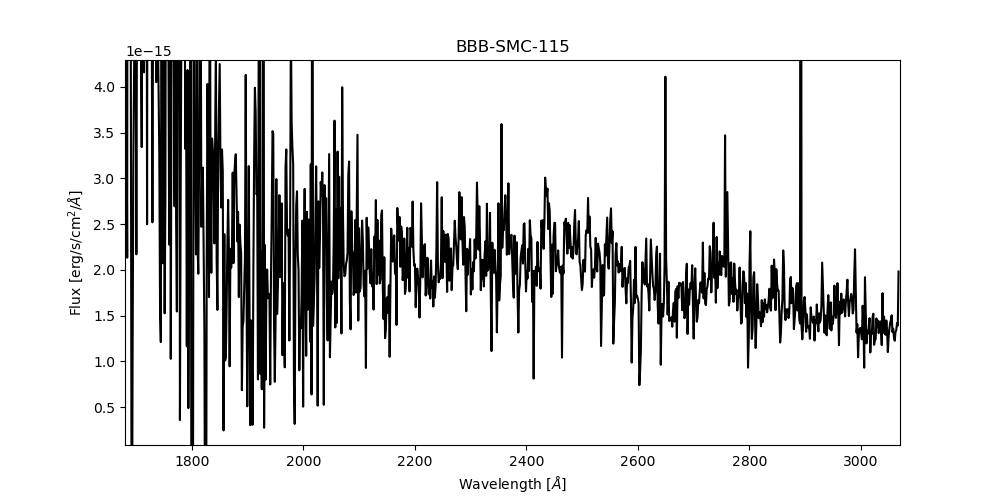}
   \caption{HST/STIS spectrum of BBB-SMC-115.}
              \label{fig:BBB-SMC-115-G230L}%
   \end{figure*}

\begin{figure*}
   \centering
    \includegraphics[width=\linewidth]{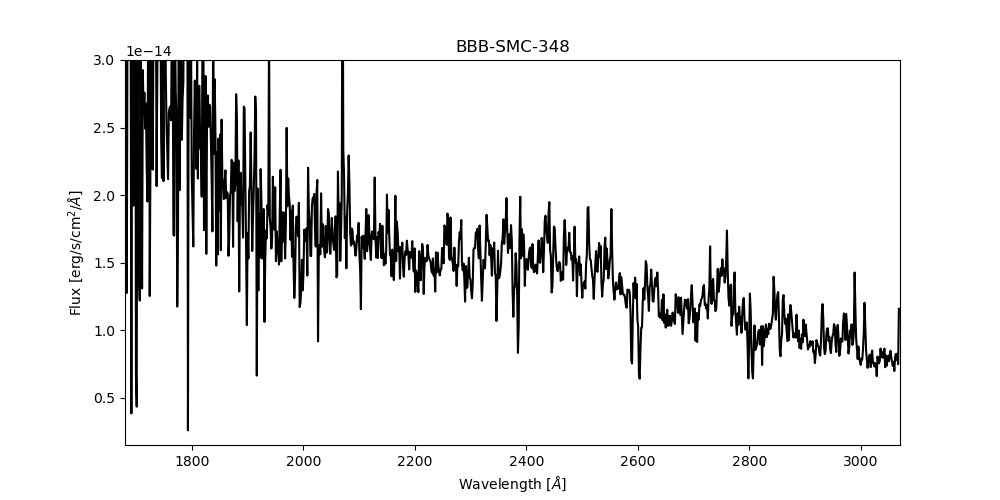}
   \caption{HST/STIS spectrum of BBB-SMC-348.}
              \label{fig:BBB-SMC-348-G230L}%
   \end{figure*}

\begin{figure*}
   \centering
    \includegraphics[width=\linewidth]{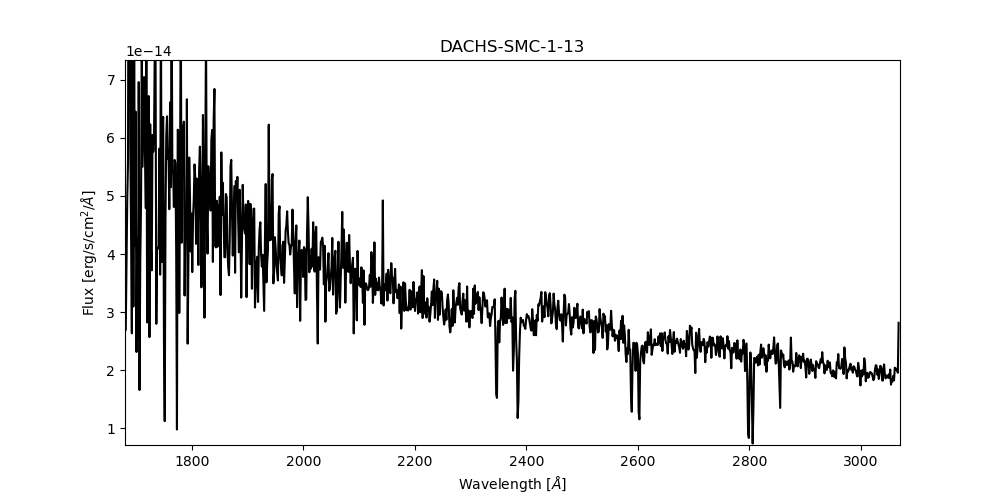}
   \caption{HST/STIS spectrum of DACHS-SMC-1-13.}
              \label{fig:DACHS-SMC-1-13-G230L}%
   \end{figure*}

\begin{figure*}
   \centering
    \includegraphics[width=\linewidth]{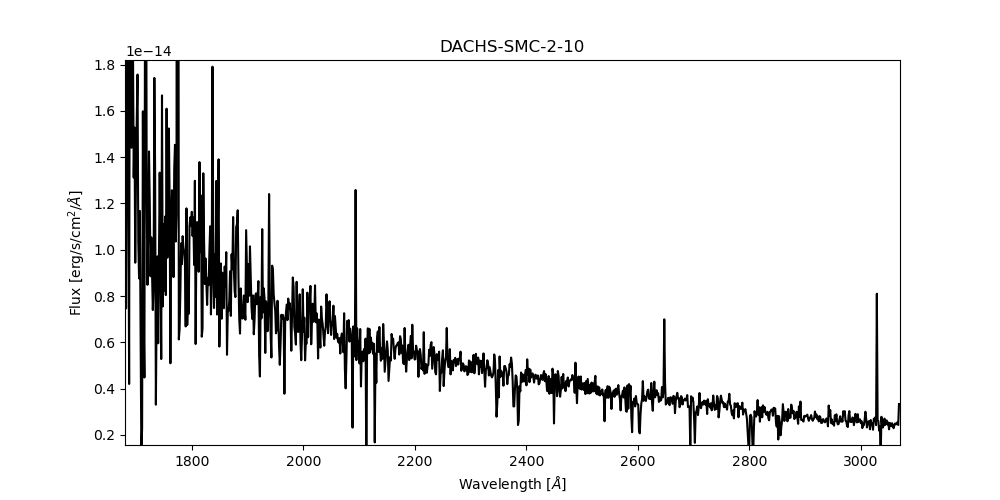}
   \caption{HST/STIS spectrum of DACHS-SMC-2-10.}
              \label{fig:DACHS-SMC-2-10-G230L}%
   \end{figure*}

\begin{figure*}
   \centering
    \includegraphics[width=\linewidth]{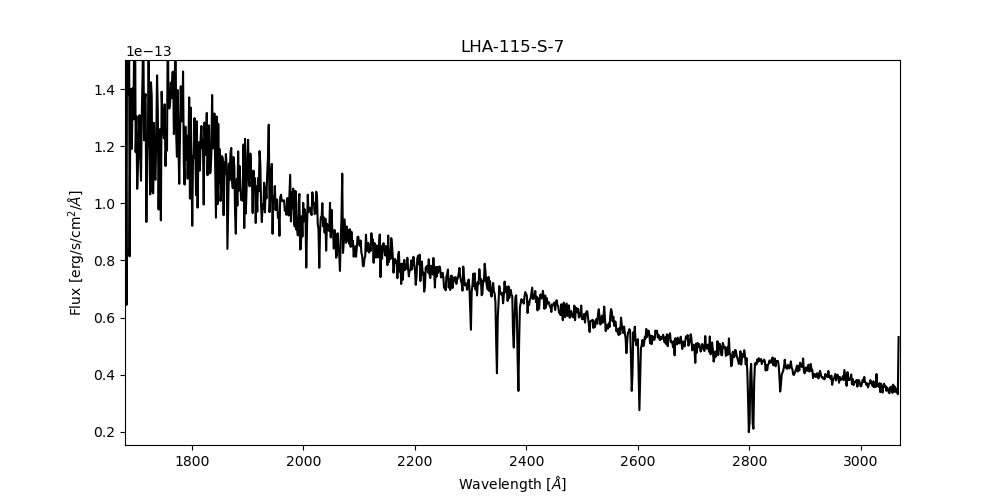}
   \caption{HST/STIS spectrum of LHA-115-S-7.}
              \label{fig:LHA-115-S-7-G230L}%
   \end{figure*}

\begin{figure*}
   \centering
    \includegraphics[width=\linewidth]{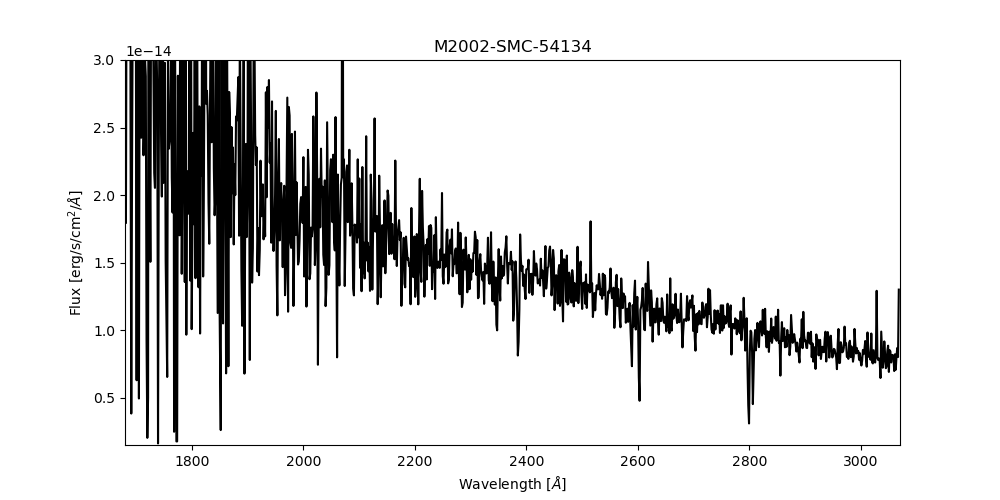}
   \caption{HST/STIS spectrum of M2002-SMC-54134.}
              \label{fig:M2002-SMC-54134-G230L}%
   \end{figure*}

\begin{figure*}
   \centering
    \includegraphics[width=\linewidth]{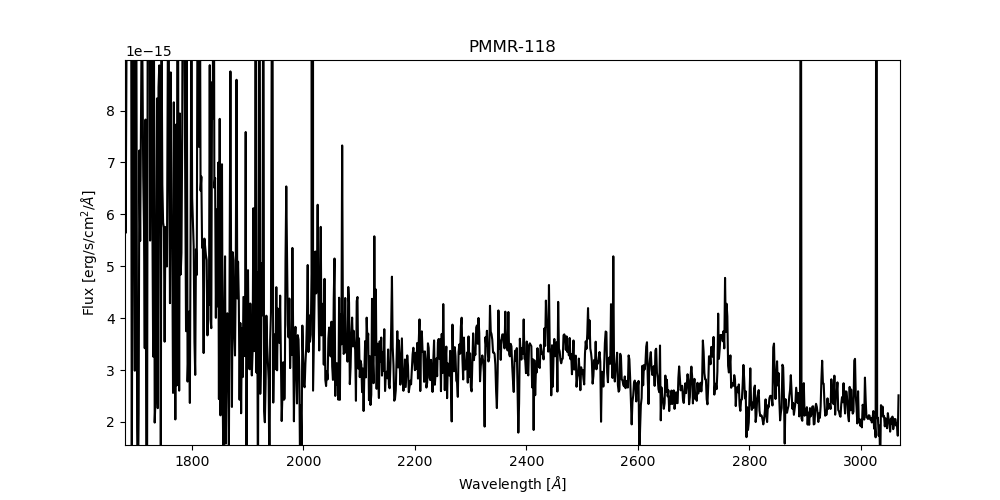}
   \caption{HST/STIS spectrum of PMMR-118.}
              \label{fig:PMMR-118-G230L}%
   \end{figure*}

\begin{figure*}
   \centering
    \includegraphics[width=\linewidth]{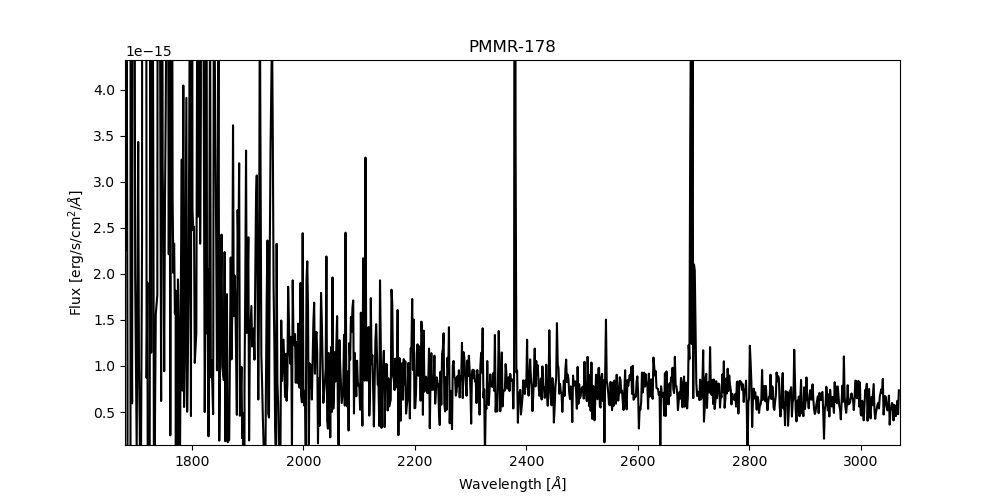}
   \caption{HST/STIS spectrum of PMMR-178.}
              \label{fig:PMMR-178-G230L}%
   \end{figure*}

\begin{figure*}
   \centering
    \includegraphics[width=\linewidth]{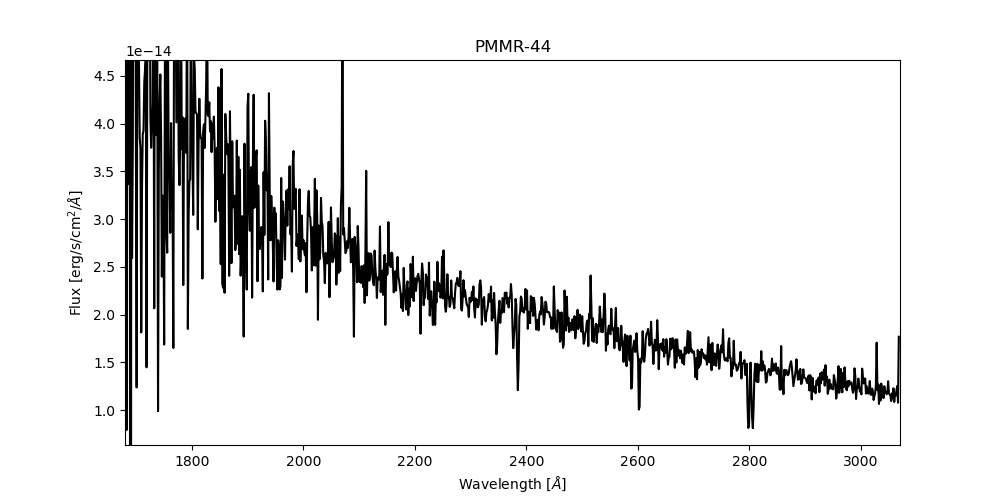}
   \caption{HST/STIS spectrum of PMMR-44.}
              \label{fig:PMMR-44-G230L}%
   \end{figure*}

\begin{figure*}
   \centering
    \includegraphics[width=\linewidth]{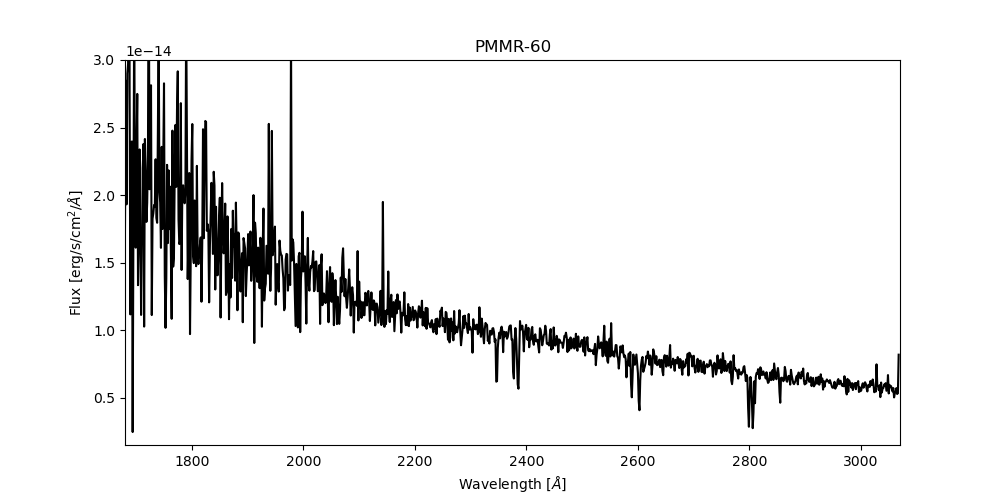}
   \caption{HST/STIS spectrum of PMMR-60.}
              \label{fig:PMMR-60-G230L}%
   \end{figure*}

\begin{figure*}
   \centering
    \includegraphics[width=\linewidth]{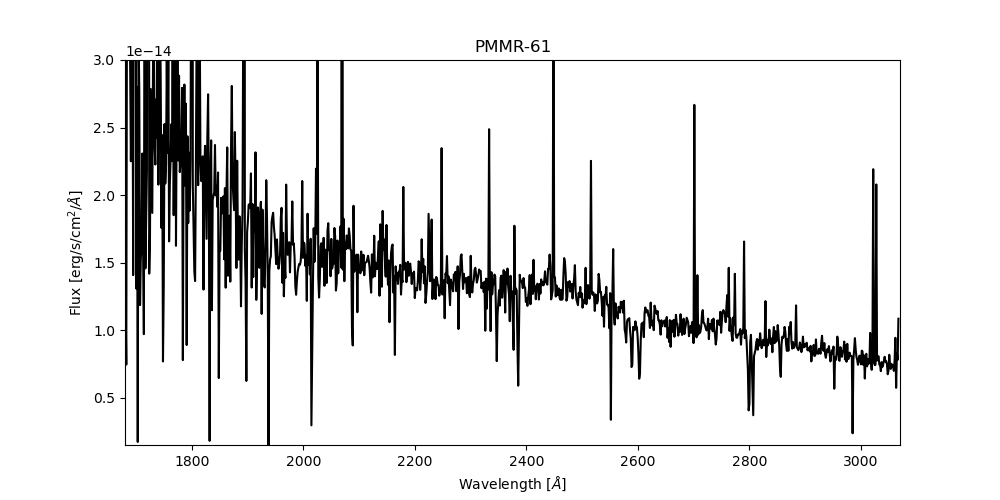}
   \caption{HST/STIS spectrum of PMMR-61.}
              \label{fig:PMMR-61-G230L}%
   \end{figure*}

\begin{figure*}
   \centering
    \includegraphics[width=\linewidth]{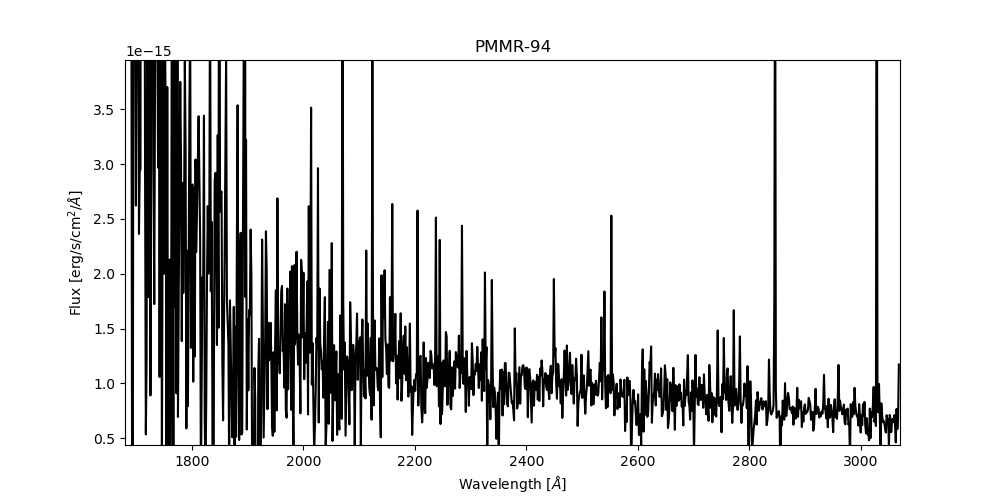}
   \caption{HST/STIS spectrum of PMMR-94.}
              \label{fig:PMMR-94-G230L}%
   \end{figure*}

\begin{figure*}
   \centering
    \includegraphics[width=\linewidth]{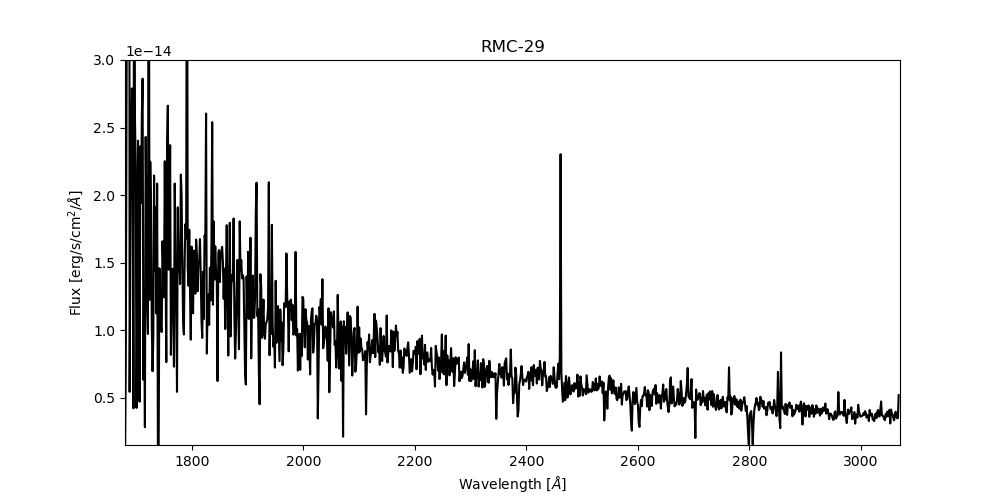}
   \caption{HST/STIS spectrum of RMC-29.}
              \label{fig:RMC-29-G230L}%
   \end{figure*}

\begin{figure*}
   \centering
    \includegraphics[width=\linewidth]{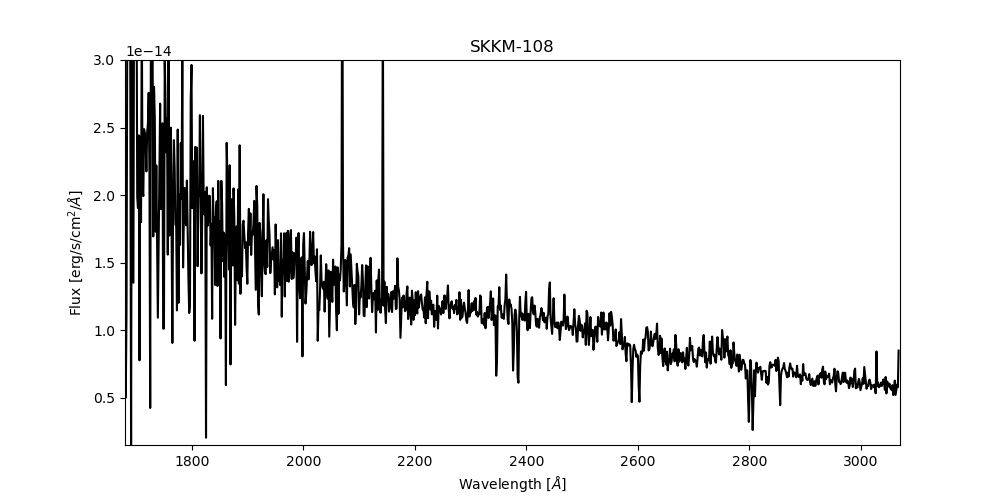}
   \caption{HST/STIS spectrum of SKKM~108.}
              \label{fig:SKKM-108-G230L}%
   \end{figure*}

\begin{figure*}
   \centering
    \includegraphics[width=\linewidth]{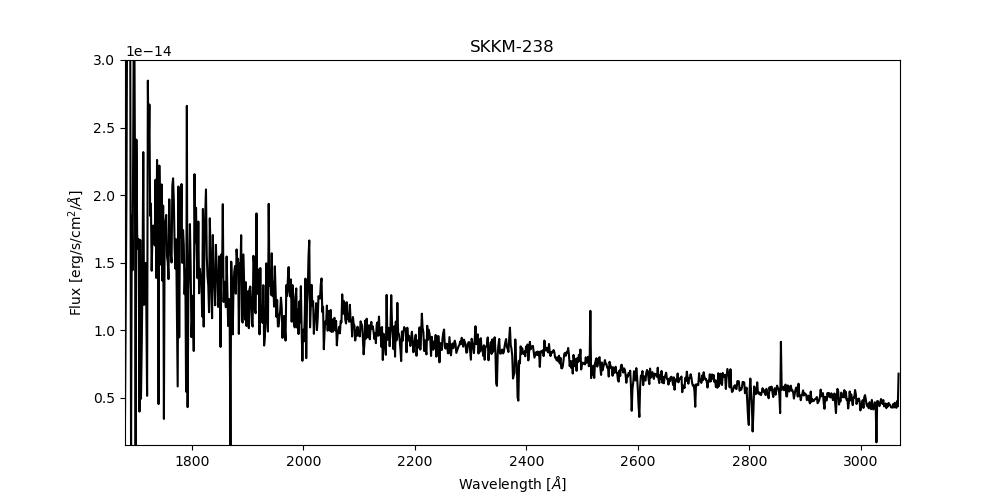}
   \caption{HST/STIS spectrum of SKKM~238.}
              \label{fig:SKKM-238-G230L}%
   \end{figure*}

\begin{figure*}
   \centering
    \includegraphics[width=\linewidth]{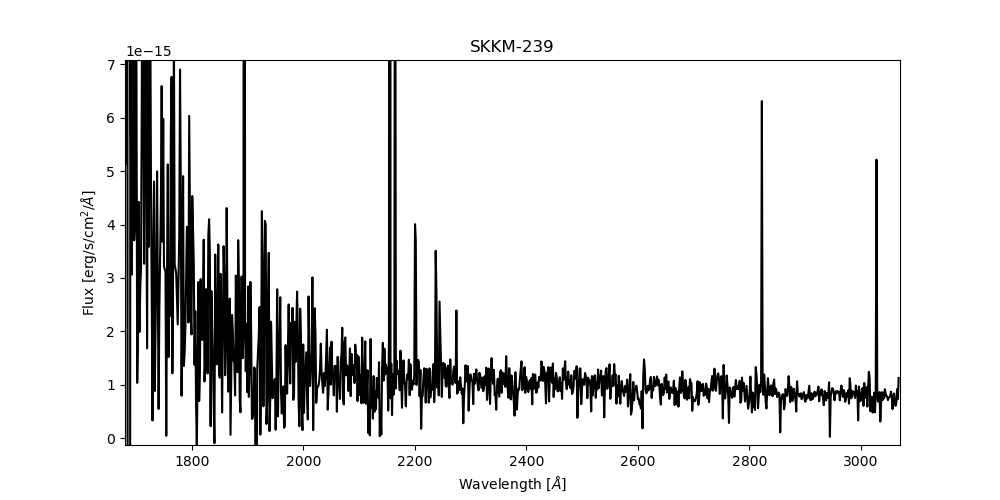}
   \caption{HST/STIS spectrum of SKKM~239.}
              \label{fig:SKKM-239-G230L}%
   \end{figure*}

Figures~\ref{fig:BBB-SMC-115-G230L} -- \ref{fig:SKKM-239-G230L} display the HST Snap spectra for all 16 targets considered in this study. 

\section{Individual extinction parameters for SMC RSGs} \label{sub:ebv}
Table~\ref{tb:ebv} displays the reddening values determined from the 20 nearest O-type stars. See Section~\ref{sec:results} for further details on this analysis.
In general, the $E(B-V)$ values are small with no examples of sources being located in regions of higher intrinsic extinction within the SMC.
This gives us confidence that the assumed extinction parameters in the fits are robust.  

\begin{table}
\caption{Table of $E(B-V)$ values determined from the 20 nearest O-type stars.}
\label{tb:ebv}
\begin{tabular}{l c}
\hline
ID                    & $E(B-V)_{\rm O-star}$ \\
\hline
SkKM 74               & 0.094 \\
SkKM 90               & 0.094 \\
SkKM 108              & 0.076 \\
SkKM 113              & 0.076 \\
SkKM 171              & 0.088 \\
SkKM 209              & 0.067 \\
SkKM 238              & 0.064 \\
SkKM 239              & 0.076 \\
SkKM 286              & 0.066 \\
SkKM 310              & 0.066 \\
SkKM 320              & 0.051 \\
$[$M2002$]$~SMC~54134 & 0.076 \\
PMMR  61              & 0.077 \\
SkKM 247              & 0.076 \\
SkKM 250              & 0.076 \\
LHA 115-S   7         & 0.117 \\
\hline
\end{tabular}
\end{table}

\begin{figure}
   \centering
   \includegraphics[width=\linewidth]{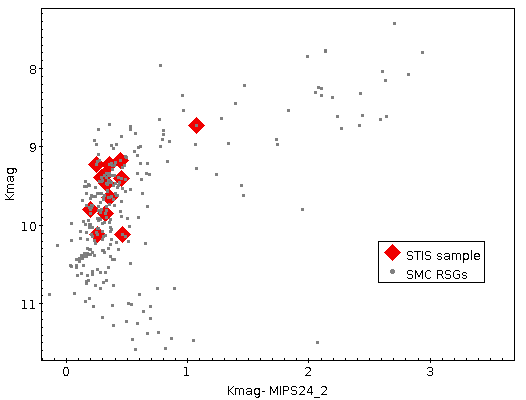}
   \caption{$Ks$-band magnitudes ($M_{Ks})$ vs $Ks - 24 \mu$m colour-magnitude diagram for SMC RSGs in grey points from Patrick et al. (in prep.).
      The red diamonds show RSGs observed with UV SITS spectroscopy.}
              \label{fig:K-24}%
    \end{figure}

The extinction parameters derived using this method are not sensitive to circumstellar extinction intrinsic to the RSG, which is sometimes reported to be up to 0.5\,mag greater than OB-stars in the same region~\citep{2009ApJ...703..420M}. 
In an attempt to asses the extinction parameters on an individual level we examine the recent results of \citet{2023ApJ...946...43W}.
\citet{2023ApJ...946...43W} determine colour excess values and $E(G_{BP}-G_{RP})$ values for several hundred SMC RSGs.
By cross-matching our target list with their catalogue and converting the $E(G_{BP}-G_{RP})$ to $E(B-V)$ using their Table~4, we find 8 sources in common with $-0.08 < E(B-V) < 0.08$. 
The negative values here indicate that the uncertainty on each individual value is at least $\Delta E(B-V) = 0.08$.
Based on this comparison we find no evidence for increased reddening values for RSGs in the SMC.
\citet{2023ApJ...946...43W} determined $R_v = 2.53$\footnote{We note here that this value is largely inconsistent with other studies. For the SMC $R_V$ is usually in the range $2.74 < R_V < 3.1$.}, therefore these targets display $A_V$ values in the range $-0.23 < A_V < 0.24$ with formal uncertainties on individual $A_V$ values in the range $0 < \delta A_V < 0.01$. 
Figure~\ref{fig:K-24} displays the $K{\rm s}-[24\mu m]$ colours of RSGs in the SMC, following~\citet{2010AJ....140..416B}.
Stars with higher dust emission and circumstellar material should display larger $K_{\rm s}-[24\mu m]$ colours.
This figure illustrates, that one target, SkKM~247, shows evidence for increased circumstellar material with respect to the typical level of circumstellar material displayed for SMC RSGs.
In contrast, SkKM~171 and SkKM~239, which are the two RSGs in the sample with the highest values of $A_v$ from \citet{2023ApJ...946...43W}, show no evidence for increased levels of circumstellar material at 24\,$\mu m$.
This likely reflects the intrinsic uncertainty on the individual extinction parameters from \citet{2023ApJ...946...43W}.
We assume this intrinsic uncertainty is a product of the variability of RSGs and the difficulty that such variability presents on the definition of an intrinsic colour as a function of effective temperature for RSGs.
Moreover, intrinsic effective temperature changes accompany variability~\citep{2019A&A...632A..28K, 2021A&A...650L..17K}, and RSGs in the SMC in particular are known to exhibit large spectral type variability~\citep{2018A&A...618A.137D} which further complicates such an analysis. 
Three of the 8 targets observed with STIS have positive, significant values of $A_v$ as determined by \citet{2023ApJ...946...43W}. 
Because of this uncertainty we adopt the reddening parameters determined from the comparison with the O-type stars.

\section{Spectral energy distribution fits} \label{sec:SEDs}
\begin{figure*}
   \centering
   \includegraphics[width=0.9\linewidth]{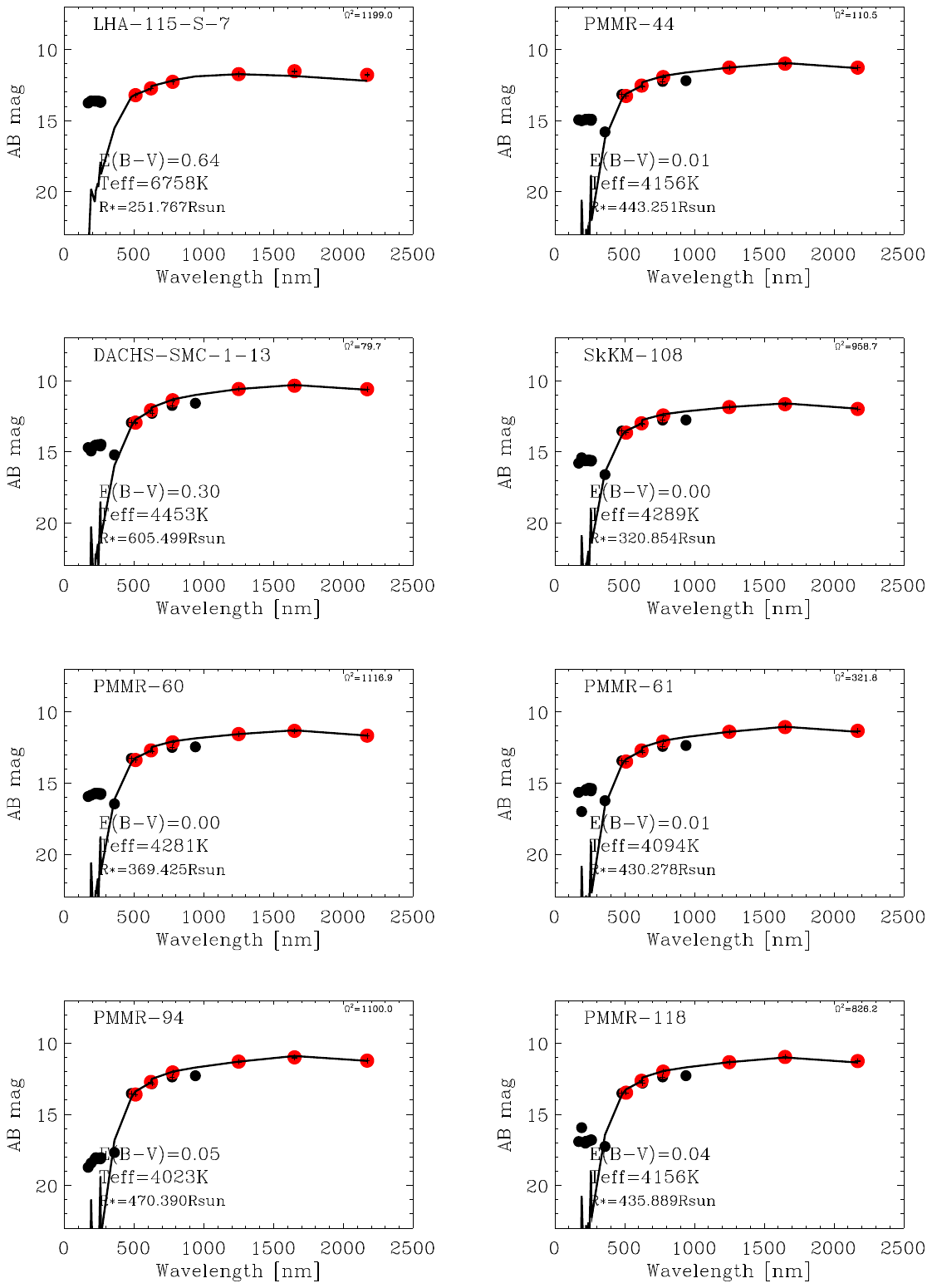}
   \caption{SED fits for all 16 targets showing the best fit Castelli \& Kurucz model (in black) and associated stellar and extinction parameters. The red points denote photometry given weight in the fit. The black points are shown for comparison.}
              \label{fig:SED1}%
    \end{figure*}

\begin{figure*}
   \centering
   \includegraphics[width=0.9\linewidth]{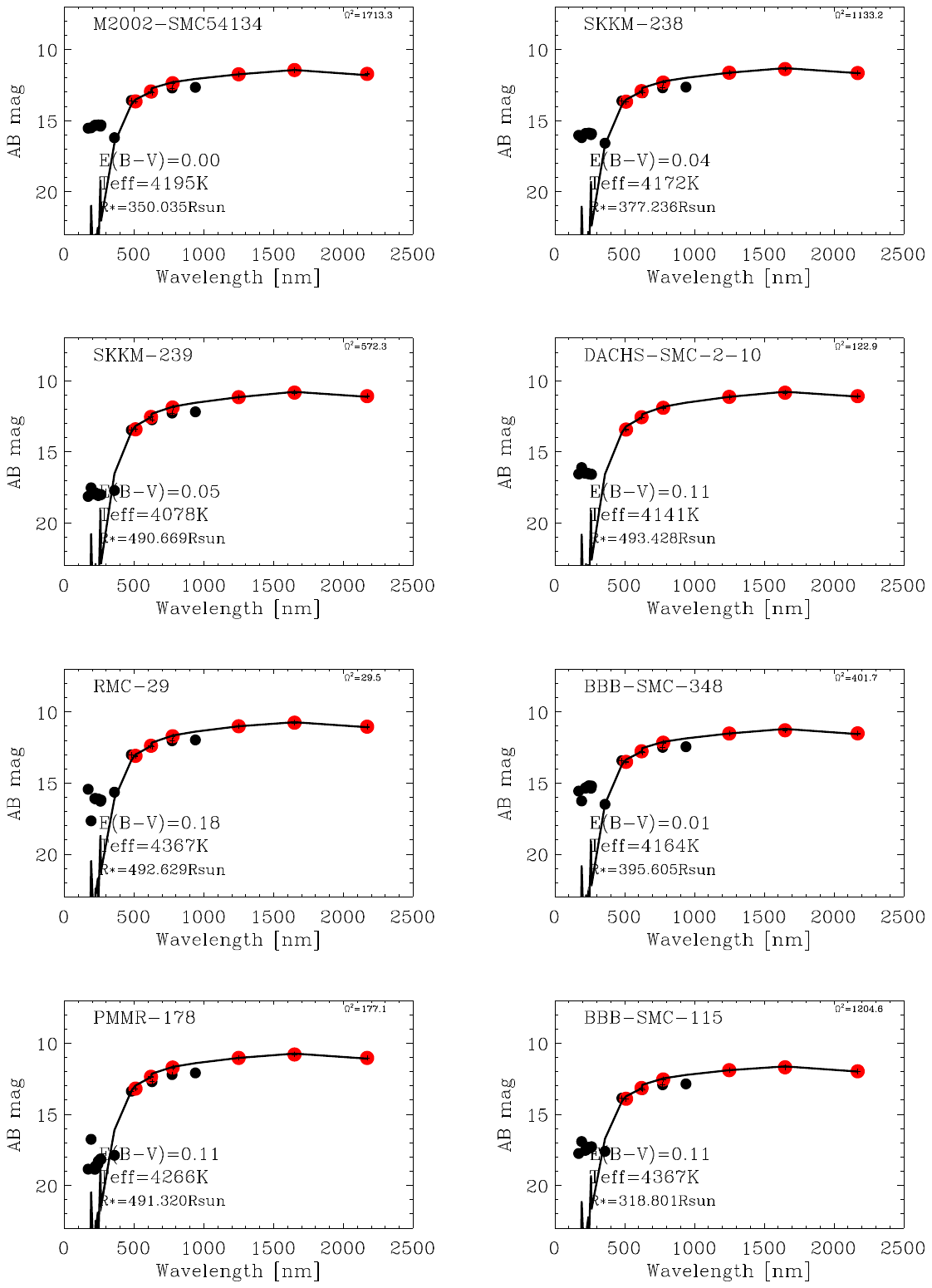}
   \caption{Figure~\ref{fig:SED1} continued.}
              \label{fig:SED2}%
    \end{figure*}

Figures~\ref{fig:SED1} and~\ref{fig:SED2} present the SED fits for all targets using the method described in Section~\ref{sub:SED}.


\section{Spectral type variability} \label{sec:spt_var}
Table~\ref{tb:RMC29} displays the various spectral classifications for the star SkKM~250.

\begin{table}
\caption{RMC 29 spectral type variability}
\label{tb:RMC29}
\begin{tabular}{lll}
\hline
ID                 & SpT & Reference\\
\hline
RMC  29  &  K2\,Ia     & {\cite{2018A&A...618A.137D}} 2010 \\
RMC  29  &  K0\,Ia-Iab & {\cite{2018A&A...618A.137D}} 2010 \\
RMC  29  &  K2\,Iab    & {\cite{2018A&A...618A.137D}} 2010 \\
RMC  29  &  K4\,Ia-Iab & {\cite{2018A&A...618A.137D}} 2012 \\
RMC  29  &  K3--5\,I     & {\cite{2003AJ....126.2867M}} \\
RMC  29  &  K4\,Ia-Iab   & {\citet{2015A&A...578A...3G}} \\
RMC  29  &  K1\,Ia-Iab   & {\citet{2015A&A...578A...3G}} \\
RMC  29  &  K5-8         & {\citet{1983A&AS...53..255P}} \\
\hline
\end{tabular}
\end{table}
\end{appendix}

\end{document}